\@twosidetrue
\@twocolumntrue
\@mparswitchtrue
\def\ds@draft{\overfullrule 5pt}
\def\ds@twocolumn{\@twocolumntrue}
\def\ds@onecolumn{\@twocolumnfalse}
\newif\ifSFB@landscape
\def\ds@landscape{\SFB@landscapetrue}
\newif\ifSFB@galley
\def\ds@galley{\SFB@galleytrue}
\newif\ifSFB@referee
\def\ds@referee{%
 \SFB@refereetrue
 \@twocolumnfalse
}
\@options
\ifSFB@referee
 
\else
 
\fi
\if@twocolumn
  \def\@normalsize{\@setsize\normalsize{11pt}\ixpt\@ixpt
   \abovedisplayskip 6pt plus 2pt minus 2pt
   \belowdisplayskip \abovedisplayskip
   \abovedisplayshortskip 6pt plus 2pt
   \belowdisplayshortskip \abovedisplayshortskip
   \let\@listi\@listI}
 \else
 \ifSFB@referee
  \def\@normalsize{\@setsize\normalsize{14pt}\xiipt\@xiipt
   \abovedisplayskip 4pt plus 1pt minus 1pt
   \belowdisplayskip \abovedisplayskip
   \abovedisplayshortskip 4pt plus 1pt
   \belowdisplayshortskip \abovedisplayshortskip
   \let\@listi\@listI}
 \else
  \def\@normalsize{\@setsize\normalsize{12pt}\ixpt\@ixpt
   \abovedisplayskip 4pt plus 1pt minus 1pt
   \belowdisplayskip \abovedisplayskip
   \abovedisplayshortskip 4pt plus 1pt
   \belowdisplayshortskip \abovedisplayshortskip
   \let\@listi\@listI}
 \fi
\fi
\def\small{\@setsize\small{10pt}\viiipt\@viiipt
 \abovedisplayskip 4pt plus 1pt minus 1pt
 \belowdisplayskip \abovedisplayskip
 \abovedisplayshortskip 4pt plus 1pt
 \belowdisplayshortskip \abovedisplayshortskip
 \def\@listi{\leftmargin\leftmargini
  \topsep 2pt plus 1pt minus 1pt
  \parsep \z@
  \itemsep 2pt}}
\def\footnotesize{\@setsize\footnotesize{10pt}\viiipt\@viiipt
 \abovedisplayskip 4pt plus 1pt minus 1pt
 \belowdisplayskip \abovedisplayskip
 \abovedisplayshortskip 4pt plus 1pt
 \belowdisplayshortskip \abovedisplayshortskip
 \def\@listi{\leftmargin\leftmargini
  \topsep 2pt plus 1pt minus 1pt
  \parsep \z@
  \itemsep 2pt}}
\def\scriptsize{\@setsize\scriptsize{8pt}\viipt\@viipt}
\def\tiny{\@setsize\tiny{6pt}\vpt\@vpt}
\if@twocolumn
  \def\large{\@setsize\large{11pt}\xpt\@xpt}
 \else
  \def\large{\@setsize\large{12pt}\xpt\@xpt}
 \fi
\def\Large{\@setsize\Large{14pt}\xiipt\@xiipt}
\def\LARGE{\@setsize\LARGE{17pt}\xivpt\@xivpt}
\def\huge{\@setsize\huge{20pt}\xviipt\@xviipt}
\def\Huge{\@setsize\huge{25pt}\xxpt\@xxpt}
\normalsize

\if@twocolumn
\else
 \ifSFB@referee
  \oddsidemargin \z@  \evensidemargin \z@
 \else
 \fi
\fi
 \topmargin \z@
\newdimen\SFB@measure
\textwidth \SFB@measure
\ifSFB@landscape
 \textwidth \textheight
 \textheight \SFB@measure
\fi
\ifSFB@referee
\fi
\skip\footins 19.5pt plus 12pt minus 1pt

\parskip \z@ plus .1pt
\@beginparpenalty -\@lowpenalty
\@endparpenalty -\@lowpenalty
\@itempenalty -\@lowpenalty
\newcounter{part}
\newcounter {section}
\newcounter {subsection}[section]
\newcounter {subsubsection}[subsection]
\newcounter {paragraph}[subsubsection]
\newcounter {subparagraph}[paragraph]
\def\thepart          {\arabic{part}}
\def\thesection       {\arabic{section}}

\def\part{\par \addvspace{4ex}\@afterindentfalse
 \secdef\@part\@spart}
\def\@part[#1]#2{\ifnum \c@secnumdepth >\m@ne
  \refstepcounter{part}
  \addcontentsline{toc}{part}{Part \thepart: #1}
 \else \addcontentsline{toc}{part}{#1}
 \fi
 {\parindent 0pt \raggedright
  \ifnum \c@secnumdepth >\m@ne
   \large\rm PART
   \ifcase\thepart \or ONE \or TWO \or THREE \or FOUR \or FIVE
    \or SIX \or SEVEN \or EIGHT \or NINE \or TEN \else \fi
   \par \nobreak
  \fi
  \LARGE \rm #2 \markboth{}{}\par }
 \nobreak \vskip 3ex \@afterheading}
\def\@spart#1{{\parindent 0pt \raggedright
  \LARGE \rm #1\par}
 \nobreak \vskip 3ex \@afterheading}

\def\section{\@startsection {section}{1}{\z@}
 {-24pt plus -12pt minus -1pt}
 {6pt}
 {\SFB@hangraggedright\normalsize\bf}}
\def\subsection{\@startsection{subsection}{2}{\z@}
 {-18pt plus -9pt minus -1pt}
 {6pt}
 {\SFB@hangraggedright\large\bf}}
\def\subsubsection{\@startsection{subsubsection}{3}{\z@}
 {-18pt plus -9pt minus -1pt}
 {6pt}
 {\SFB@hangraggedright\normalsize\it}}
\def\paragraph{\@startsection{paragraph}{4}{\z@}
 {12pt plus 2.25pt minus 1pt}{-0.5em}{\normalsize\bf}}
\def\subparagraph{\@startsection{subparagraph}{5}{\parindent}
 {12pt plus 2.25pt minus 1pt}{-0.5em}{\normalsize\it}}
\def\SFB@hangraggedright{\rightskip\@flushglue \let\\=\newline}
\def\@sect#1#2#3#4#5#6[#7]#8{%
 \ifnum #2>\c@secnumdepth
  \def\@svsec{}%
 \else
  \refstepcounter{#1}
  \ifnum #2=\@ne
   \ifSFB@appendix \edef\@svsec{}%
             \else \edef\@svsec{\csname the#1\endcsname\hskip 1em}%
   \fi
  \else \edef\@svsec{\csname the#1\endcsname\hskip 1em}%
  \fi
 \fi
 \@tempskipa #5\relax
 \ifdim \@tempskipa>\z@
  \begingroup #6\relax
   \ifnum #2=\@ne
    \ifSFB@appendix
     \@hangfrom{\hskip #3\relax\@svsec}{\interlinepenalty \@M
      APPENDIX \csname the#1\endcsname:\hskip 0.5em\uppercase{#8}\par}%
    \else
     \@hangfrom{\hskip #3\relax\@svsec}{\interlinepenalty \@M
      \uppercase{#8}\par}%
    \fi
   \else
    \@hangfrom{\hskip #3\relax\@svsec}{\interlinepenalty \@M #8\par}%
   \fi
  \endgroup
  \csname #1mark\endcsname{#7}%
  \addcontentsline{toc}{#1}{\ifnum #2>\c@secnumdepth \else
   \protect\numberline{\csname the#1\endcsname}\fi #7}%
 \else
  \def\@svsechd{#6\hskip #3\@svsec \ifnum #2=\@ne\uppercase{#8}\else #8\fi
  \csname #1mark\endcsname{#7}
  \addcontentsline{toc}{#1}{\ifnum #2>\c@secnumdepth \else
   \protect\numberline{\csname the#1\endcsname}\fi#7}}%
 \fi
 \@xsect{#5}}

\newif\ifSFB@appendix
\def\appendix{\par
 \SFB@appendixtrue
 \setcounter{section}{0}
 \def\thesection{A\arabic{section}}
 \setcounter{equation}{0}
 \def\theequation{A\arabic{equation}}
 \setcounter{figure}{0}
 \def\thefigure{A\@arabic\c@figure}
 \setcounter{table}{0}
 \def\thetable{A\@arabic\c@table}
}

\newskip\@indentskip
\newskip\smallindent
\newskip\@footindent
\newskip\@leftskip
\@footindent=\smallindent
\@leftskip=\z@

\leftmargini   \@indentskip
\leftmargin\leftmargini
\labelwidth\leftmargini\advance\labelwidth-\labelsep
\def\makeRRlabel#1{\hss\llap{#1}}
\def\@listI{\leftmargin\leftmargini
 \parsep \z@
 \topsep 6pt plus 1pt minus 1pt
 \itemsep \z@ plus .1pt
}
\let\@listi\@listI
\@listi
\def\@listii{\leftmargin\leftmarginii
 \labelwidth\leftmarginii\advance\labelwidth-\labelsep
 \topsep 6pt plus 1pt minus 1pt
 \parsep \z@
 \itemsep \z@ plus .1pt
}
\def\@listiii{\leftmargin\leftmarginiii
 \labelwidth\leftmarginiii\advance\labelwidth-\labelsep
 \topsep 6pt plus 1pt minus 1pt
 \parsep \z@
 \partopsep \z@
 \itemsep \topsep
}
\def\@listiv{\leftmargin\leftmarginiv
 \labelwidth\leftmarginiv\advance\labelwidth-\labelsep
}
\def\@listv{\leftmargin\leftmarginv
 \labelwidth\leftmarginv\advance\labelwidth-\labelsep
}
\def\@listvi{\leftmargin\leftmarginvi
 \labelwidth\leftmarginvi\advance\labelwidth-\labelsep
}
\def\itemize{\ifnum \@itemdepth >3 \@toodeep
  \else \advance\@itemdepth \@ne
   \edef\@itemitem{labelitem\romannumeral\the\@itemdepth}%
   \list{\csname\@itemitem\endcsname}%
    {\let\makelabel\makeRRlabel}%
  \fi}

\def\enumerate{\ifnum \@enumdepth >3 \@toodeep \else
  \advance\@enumdepth \@ne
  \edef\@enumctr{enum\romannumeral\the\@enumdepth}%
 \fi
 \@ifnextchar [{\@enumeratetwo}{\@enumerateone}%
}
\def\@enumeratetwo[#1]{%
 \list{\csname label\@enumctr\endcsname}%
  {\settowidth\labelwidth{[#1]}
   \leftmargin\labelwidth \advance\leftmargin\labelsep
   \usecounter{\@enumctr}
   \let\makelabel\makeRRlabel}
}
\def\@enumerateone{%
 \list{\csname label\@enumctr\endcsname}%
  {\usecounter{\@enumctr}
   \let\makelabel\makeRRlabel}}

\def\theenumi{(\roman{enumi})}

\def\theenumii{(\alph{enumii})}
\def\p@enumii{\theenumi}

\def\theenumiii{(\arabic{enumiii})}
\def\p@enumiii{\theenumi(\theenumii)}

\def\p@enumiv{\p@enumiii\theenumiii}
\def\description{\list{}{\labelwidth\z@ \itemindent-\leftmargin
  \leftmargin 1em
  \itemindent-1em
}}

\def\verse{\let\\=\@centercr
 \list{}{\itemsep\z@
  \itemindent -\@indentskip
  \listparindent \itemindent
  \rightmargin\leftmargin
  \advance\leftmargin \@indentskip}\item[]}

\def\quotation{\list{}{\listparindent \smallindent
  \leftmargin\z@\rightmargin\leftmargin
  \parsep 0pt plus 1pt}\item[]\small}

\def\quote{\list{}{\leftmargin\z@\rightmargin\leftmargin}\item[]\small}

\def\@begintheorem#1#2{\rm \trivlist \item[\hskip \labelsep{\bf #1\ #2.}]}
\def\@opargbegintheorem#1#2#3{\rm \trivlist
  \item[\hskip \labelsep{\bf #1\ #2.\ (#3)}]}
\def\@endtheorem{\endtrivlist}
\@namedef{proof*}{\rm \trivlist \item[\hskip \labelsep{\it Proof.}]}
\@namedef{endproof*}{\endtrivlist}

\def\titlepage{\@restonecolfalse\if@twocolumn\@restonecoltrue\onecolumn
  \else \newpage \fi \thispagestyle{empty}\c@page\z@}
\def\endtitlepage{\if@restonecol\twocolumn \else \newpage \fi}

\def\tabular{\def\@halignto{}
 \def\hline{\noalign{\ifnum0=`}\fi
  \vskip 3pt
  \hrule \@height \arrayrulewidth
  \vskip 3pt
  \futurelet \@tempa\@xhline}
 \def\fullhline{\noalign{\ifnum0=`}\fi
  \vskip 3pt
  \hrule \@height \arrayrulewidth
  \vskip 3pt
  \futurelet \@tempa\@xhline}
 \def\@xhline{\ifx\@tempa\hline
   \vskip -6pt
   \vskip \doublerulesep
  \fi
  \ifnum0=`{\fi}}
  \def\@arrayrule{\@addtopreamble{\hskip -.5\arrayrulewidth
                                  \hskip .5\arrayrulewidth}}
\@tabular
}
\tabbingsep \labelsep

\skip\@mpfootins = \skip\footins

\fboxrule = \arrayrulewidth

\def\maketitle{\par
 \begingroup
  \if@twocolumn
   \twocolumn[\vspace*{17pt}\@maketitle]
  \else
   \newpage
   \global\@topnum\z@
   \@maketitle
  \fi
  \thispagestyle{titlepage}
 \endgroup
 \let\maketitle\relax
 \let\@maketitle\relax
 \gdef\@author{}
 \gdef\@title{}
 \let\thanks\relax
}
\def\and{\end{author@tabular}\vskip 6pt\par
 \begin{author@tabular}[t]{@{}l@{}}}
\def\@maketitle{\newpage
 \vspace*{7pt}
 {\raggedright \sloppy
  {\huge \bf \@title \par}
  \vskip 23pt
  {\LARGE
   \begin{author@tabular}[t]{@{}l@{}}\@author
   \end{author@tabular}\par}
  \vskip 22pt
 }
 \par\noindent
 {\small \@date \par}
 \vskip 22pt
}
\def\abstract{\if@twocolumn
  \start@SFBbox\@abstract
 \else
  \@abstract
 \fi}
\def\endabstract{\if@twocolumn
   \endlist\finish@SFBbox
 \else
  \endlist
 \fi}
\def\@abstract{\list{}{\leftmargin 10.5pc\rightmargin\z@
  \parsep 0pt plus 1pt}\item[]\normalsize{\bf ABSTRACT}\\\large} 
\newif\ifSFB@keywords
\def\keywords{\if@twocolumn
  \start@SFBbox\@keywords
 \else
  \@keywords
 \fi
}
\def\@keywords{\list{}{\leftmargin 10.5pc\rightmargin\z@
  \parsep 0pt plus 1pt}\item[]\large{\bf Key words: }}
\def\endkeywords{\if@twocolumn
  \endlist\addvspace{37pt}\finish@SFBbox
 \else
  \endlist
 \fi
 \@thanks
 \gdef\@thanks{}
 \SFB@keywordstrue
}
\def\nokeywords{\ifSFB@keywords\else
 \if@twocolumn \start@SFBbox\addvspace{37pt}\finish@SFBbox \fi
 \@thanks
 \gdef\@thanks{}\fi
}

\def\author@tabular{\def\@halignto{}\@authortable}
\let\endauthor@tabular=\endtabular
\def\author@tabcrone{{\ifnum0=`}\fi\@xtabularcr[-7pt]\small\it
 \let\\=\author@tabcrtwo\ignorespaces}
\def\author@tabcrtwo{{\ifnum0=`}\fi\@xtabularcr[-7pt]\small\it
 \let\\=\author@tabcrtwo\ignorespaces}
\def\@authortable{\leavevmode \hbox \bgroup $\let\@acol\@tabacol
 \let\@classz\@tabclassz \let\@classiv\@tabclassiv
 \let\\=\author@tabcrone \ignorespaces \@tabarray}

\def\start@SFBbox{\@next\@currbox\@freelist{}{}%
 \global\setbox\@currbox
 \vbox\bgroup
  \hsize \textwidth
  \@parboxrestore
}
\def\finish@SFBbox{\par\vskip -\dbltextfloatsep
  \egroup
  \global\count\@currbox\tw@
  \global\@dbltopnum\@ne
  \global\@dbltoproom\maxdimen\@addtodblcol
  \global\vsize\@colht
  \global\@colroom\@colht
}

\mark{{}{}}
\gdef\@author{\mbox{}}
\def\author{\@ifnextchar [{\@authortwo}{\@authorone}}
\def\@authortwo[#1]#2{\gdef\@author{#2}\gdef\@shortauthor{#1}}
\def\@authorone#1{\gdef\@author{#1}\gdef\@shortauthor{#1}}
\gdef\@shortauthor{}
\gdef\@title{\mbox{}}
\def\title{\@ifnextchar [{\@titletwo}{\@titleone}}
\def\@titletwo[#1]#2{\gdef\@title{#2}\gdef\@shorttitle{#1}}
\def\@titleone#1{\gdef\@title{#1}\gdef\@shorttitle{#1}}
\gdef\@shorttitle{}
\def\volume#1{\gdef\@volume{#1}}
\gdef\@volume{000}
\def\microfiche#1{\gdef\@microfiche{#1}}
\gdef\@microfiche{}
\def\pagerange#1{\gdef\@pagerange{#1}}
\gdef\@pagerange{000--000}
\def\journal#1{\gdef\@journal{#1}}
\gdef\@journal{{Mon.\ Not.\ R.\ Astron.\ Soc.} {\bf \@volume}, \@pagerange\
  (\number\year) \@microfiche}
\def\ps@headings{\let\@mkboth\markboth
 \def\@oddhead{\Large \hfill \it \@shorttitle \hspace{1.5em}\rm \thepage}
 \def\@oddfoot{}
 \def\@evenhead{\Large \thepage \hspace{1.5em}\it \@shortauthor \hfill}
 \def\@evenfoot{}
 \def\sectionmark##1{\markboth{##1}{}}
 \def\subsectionmark##1{\markright{##1}}}
\def\ps@myheadings{\let\@mkboth\@gobbletwo
 \def\@oddhead{\Large \it \rightmark \hfill \rm \thepage}
 \def\@oddfoot{}
 \def\@evenhead{\Large \it \leftmark \hfill \rm \thepage}
 \def\@evenfoot{}
 \def\sectionmark##1{}
 \def\subsectionmark##1{}}
\def\ps@titlepage{\let\@mkboth\@gobbletwo
 \def\@oddhead{\footnotesize\@journal\hfill}
 \def\@oddfoot{}
 \def\@evenhead{\footnotesize\@journal\hfill}
 \def\@evenfoot{}
 \def\sectionmark##1{}
 \def\subsectionmark##1{}}

\def\@pnumwidth{1.55em}
\def\@tocrmarg {2.55em}
\def\@dotsep{4.5}
\def\@undottedtocline#1#2#3#4#5{\ifnum #1>\c@tocdepth
 \else
  \vskip \z@ plus .2pt
  {\hangindent #2\relax
   \rightskip \@tocrmarg \parfillskip -\rightskip
   \parindent #2\relax \@afterindenttrue
   \interlinepenalty\@M \leavevmode
   \@tempdima #3\relax #4\nobreak \hfill \nobreak
   \hbox to\@pnumwidth{\hfil\rm \ }\par}\fi}
\def\tableofcontents{\@restonecolfalse
 \if@twocolumn\@restonecoltrue\onecolumn\fi
 \section*{CONTENTS} \@starttoc{toc}
 \if@restonecol\twocolumn\fi \par\vspace{12pt}}
\def\l@part#1#2{\addpenalty{-\@highpenalty}
 \addvspace{2.25em plus 1pt}
 \begingroup
  \parindent \z@ \rightskip \@pnumwidth
  \parfillskip -\@pnumwidth
  {\normalsize\rm
   \leavevmode \hspace*{3pc}
   #1\hfil \hbox to\@pnumwidth{\hss \ }}\par
   \nobreak \global\@nobreaktrue
   \everypar{\global\@nobreakfalse\everypar{}}\endgroup}
\def\l@section#1#2{\addpenalty{\@secpenalty}
 \@tempdima 1.5em
 \begingroup
  \parindent \z@ \rightskip \@pnumwidth
  \parfillskip -\@pnumwidth \rm \leavevmode
  \advance\leftskip\@tempdima \hskip -\leftskip
  #1\nobreak\hfil \nobreak\hbox to\@pnumwidth{\hss \ }\par
 \endgroup}
\def\l@subsection{\@undottedtocline{2}{1.5em}{2.3em}}
\def\l@subsubsection{\@undottedtocline{3}{3.8em}{3.2em}}
\def\l@paragraph{\@undottedtocline{4}{7.0em}{4.1em}}
\def\l@subparagraph{\@undottedtocline{5}{10em}{5em}}
\def\listoffigures{\@restonecolfalse
 \if@twocolumn\@restonecoltrue\onecolumn\fi
 \section*{LIST OF FIGURES\@mkboth{LIST OF FIGURES}{LIST OF FIGURES}}
 \@starttoc{lof} \if@restonecol\twocolumn\fi}
\def\l@figure{\@undottedtocline{1}{1.5em}{2.3em}}
\def\listoftables{\@restonecolfalse
 \if@twocolumn\@restonecoltrue\onecolumn\fi
 \section*{LIST OF TABLES\@mkboth{LIST OF TABLES}{LIST OF TABLES}}
 \@starttoc{lot} \if@restonecol\twocolumn\fi}
\let\l@table\l@figure

\def\thebibliography#1{\section*{REFERENCES}
 \addcontentsline{toc}{section}{REFERENCES}
 \list{}{\labelwidth\z@
         \leftmargin 1.5em
	 \itemsep \z@
	 \itemindent-\leftmargin}
 \small\raggedright
 \parindent\z@
 \parskip\z@ plus .1pt\relax
 \def\newblock{\hskip .11em plus .33em minus .07em}
 \sloppy\clubpenalty4000\widowpenalty4000
 \sfcode`\.=1000\relax
}

\def\@biblabel#1{\hspace*{\labelsep}[#1]}

\newif\if@restonecol
\def\theindex{\section*{INDEX}
 \addcontentsline{toc}{section}{INDEX}
 \footnotesize \parindent\z@ \parskip\z@ plus .1pt\relax
 \let\item\@idxitem}
\def\@idxitem{\par\hangindent 1em}

\def\endtheindex{\if@restonecol\onecolumn\else\clearpage\fi}

\def\footnoterule{\kern-3\p@ \hrule width 12pc height \z@ \kern 3\p@}

\def\@fnsymbol#1{\ifcase#1\or \mbox{$\star$}\or \dagger\or \ddagger\or
   \S \or \P \or \|\or **\or \dagger\dagger
   \or \ddagger\ddagger\or \S\S\or \P\P\or \|\|\else ***
   \fi\relax}

\long\def\@makefntext#1{\parindent 1em\noindent
  $^{\@thefnmark}$\hspace{4pt}#1}

\newcounter{table}
\def\thetable{\@arabic\c@table}
\def\fps@table{tbp}
\def\ftype@table{1}
\def\ext@table{lot}
\def\fnum@table{Table \thetable}
\def\table{\let\@makecaption=\SFB@maketablecaption\@float{table}}
\let\endtable\end@float
\@namedef{table*}{\let\@makecaption=\SFB@maketablecaption\@dblfloat{table}}
\@namedef{endtable*}{\end@dblfloat}

\newcounter{figure}
\def\thefigure{\@arabic\c@figure}
\def\fps@figure{tbp}
\def\ftype@figure{2}
\def\ext@figure{lof}
\def\fnum@figure{Figure \thefigure}
\def\figure{\let\@makecaption=\SFB@makefigurecaption\@float{figure}}
\let\endfigure\end@float
\@namedef{figure*}{\let\@makecaption=\SFB@makefigurecaption\@dblfloat{figure}}
\@namedef{endfigure*}{\end@dblfloat}

\long\def\SFB@makefigurecaption#1#2{\vskip 6pt
 \setbox\@tempboxa\hbox{\small{\bf #1.} #2}
 \ifdim \wd\@tempboxa >\hsize
  \small{\bf #1.} #2\par
 \else
  \hbox to\hsize{\hfil\box\@tempboxa\hfil}
 \fi
 \vskip 6pt
}
\long\def\SFB@maketablecaption#1#2{\vskip 6pt
 \setbox\@tempboxa\hbox{\small{\bf #1.} #2}
 \ifdim \wd\@tempboxa >\hsize
  \small{\bf #1.} #2\par
 \else
  \hbox to\hsize{\box\@tempboxa\hfill}
 \fi
 \vskip 6pt
}

\def\caption{\@ifstar{\SFB@caption\@captype}%
 {\refstepcounter\@captype \@dblarg{\@caption\@captype}}%
}
\long\def\SFB@caption#1#2{
 \begingroup
  \@parboxrestore
  \normalsize
  \@makecaption{\csname fnum@#1\endcsname}{\ignorespaces #2}\par
 \endgroup}

\def\@cite#1#2{(#1\if@tempswa , #2\fi)}
\def\@biblabel#1{}

\newlength{\bibhang}
\def\@citex[#1]#2{\if@filesw\immediate\write\@auxout{\string\citation{#2}}\fi
  \def\@citea{}\@cite{\@for\@citeb:=#2\do
    {\@citea\def\@citea{; }\@ifundefined
       {b@\@citeb}{{\bf ?}\@warning
       {Citation `\@citeb' on page \thepage \space undefined}}%
{\csname b@\@citeb\endcsname}}}{#1}}
\let\@internalcite\cite
\def\cite{\def\citename##1{##1}\@internalcite}
\def\shortcite{\def\citename##1{}\@internalcite}

\def\[{\relax\ifmmode\@badmath\else\begin{trivlist}\item[]\leavevmode
  \hbox to\linewidth\bgroup$
  \displaystyle
  \hskip\mathindent\bgroup\fi}

\def\]{\relax\ifmmode \egroup $\hfil
       \egroup \end{trivlist}\else \@badmath \fi}

\def\equation{\refstepcounter{equation}\trivlist \item[]\leavevmode
  \hbox to\linewidth\bgroup $
  \displaystyle
\hskip\mathindent}

\def\endequation{$\hfil
           \displaywidth\linewidth\@eqnnum\egroup \endtrivlist}

\def\eqnarray{\stepcounter{equation}\let\@currentlabel=\theequation
\global\@eqnswtrue
\global\@eqcnt\z@\tabskip\mathindent\let\\=\@eqncr
\abovedisplayskip\topsep\ifvmode\advance\abovedisplayskip\partopsep\fi
\belowdisplayskip\abovedisplayskip
\belowdisplayshortskip\abovedisplayskip
\abovedisplayshortskip\abovedisplayskip
$$\halign
to \linewidth\bgroup\@eqnsel\hskip\@centering$\displaystyle\tabskip\z@
  {##}$&\global\@eqcnt\@ne \hskip 2\arraycolsep \hfil${##}$\hfil
  &\global\@eqcnt\tw@ \hskip 2\arraycolsep $\displaystyle{##}$\hfil
   \tabskip\@centering&\llap{##}\tabskip\z@\cr}

\def\endeqnarray{\@@eqncr\egroup
 \global\advance\c@equation\m@ne$$\global\@ignoretrue}

\newdimen\mathindent
\mathindent = \z@

\def\today{\number\day\ \ifcase\month\or
  January\or February\or March\or April\or May\or June\or
  July\or August\or September\or October\or November\or December
 \fi \ \number\year}

\ps@headings
\ifSFB@galley
 \ps@empty
\fi
\ifSFB@referee
\fi
\if@twocolumn
 \twocolumn
\else
 \onecolumn
\fi
\documentstyle{mn}
\newcommand{\be}{\begin{equation}}
\newcommand{\ee}{\end{equation}}
\newcommand{\ba}{\begin{eqnarray}}
\newcommand{\ea}{\end{eqnarray}}
\newcommand{\bfig}{\begin{figure}}
\newcommand{\efig}{\end{figure}}
\newcommand{\et}{{\em et al. }}
\newcommand{\ie}{{\em ie. }}
\newcommand{\rgl}{\rangle}
\newcommand{\lgl}{\langle}
\newcommand{\x}{{\bf x}}

\newcommand{\k}{{\bf k}}

\newcommand{\q}{{\bf q}}

\newcommand{\r}{{\bf r}}

\newcommand{\n}{{\bf n}}
\newcommand{\hx}{\hat{x}}

\newcommand{\hxb}{\hat{\x}}

\newcommand{\vb}{{\bf v}}
\newcommand{\pot}{\varphi}
\newcommand{\txb}{\tilde{\x}}
\newcommand{\tx}{\tilde{x}}
\newcommand{\tJ}{\tilde{J}}

\newcommand{\trho}{\tilde{\rho}}
\newcommand{\tdelta}{\tilde{\delta}}
\newcommand{\tDelta}{\tilde{\Delta}}

\newcommand{\tb}{\tau}
\newcommand{\xt}{\x,t}
\newcommand{\zz}{\q,\tb}

\newcommand{\de}{\partial}

\newcommand{\idaa}{\frac{a}{\dot{a}}}
\newcommand{\dxj}{\frac{\de}{\de x^j}}

\newcommand{\xib}{\xi}

\newcommand{\omegab}{\omega}
\newcommand{\nab}{\nabla}

\newcommand{\e}{\equiv}
\newcommand{\ssim}{\approx}
\newcommand{\nn}{\nonumber \\}
\newcommand{\lfe}{\lefteqn}

\newcommand{\ls}{\raisebox{-.8ex}{$\buildrel{\textstyle<}\over\sim$}}
\newcommand{\gs}{\raisebox{-.8ex}{$\buildrel{\textstyle>}\over\sim$}}
\newcommand{\nd}{\noindent}

\newcommand{\spss}{\hspace{1 cm}}
\newcommand{\apj}{{\sl Astrophys. J.}}
\newcommand{\apjl}{{\sl Astrophys. J. Lett.}}

\newcommand{\mn}{{\sl Mon. Not. R. astr. Soc.}}
\newcommand{\na}{{\sl Nature}}

\newcommand{\asap}{{\sl Astr. Astrophys.}}
\newtheorem{figudumx}{Figure}
\newenvironment{figdumy}{\begin{figudumx}\rm}{\end{figudumx}}
\newcommand{\bfg}{\begin{figdumy}}
\newcommand{\efg}{\end{figdumy}}
\font\japit = cmti10 at 11truept
\title[Reconstruction Analysis]
{
\vglue-3.0truecm
\centerline{\japit Accepted for publication in Monthly Notices of the R.A.S.}
\vglue 2.5truecm
\noindent
Reconstruction Analysis I. \\
	\vspace{0.3in}
	Redshift--space deprojection in
 	the quasi--nonlinear regime
}

\author[ A.N. Taylor and M. Rowan--Robinson]
{ A. N. Taylor\thanks{Present address: Department of
 	Astronomy, University of Edinburgh, Royal Observatory, Blackford Hill,
 	Edinburgh, EH9 3HJ.} and M. Rowan--Robinson \\
        Theoretical Astronomy Unit, School of Mathematical
 	Sciences,  Queen Mary and Westfield College,Mile End Road,
        London, E1 4NS.}

\begin{document}

\maketitle

\begin{abstract}

	We present a new method for extracting the true 3-d velocity and
density fields from the nonlinear redshift--space projected density field. The
method is based on the nonlinear, nonlocal transformation of the density field.
We assume a curl--free velocity field, although in the formulation presented
here this assumption can be dropped. Toy models with special symmetry are used
to test the intrinsic accuracy of the method compared to linear theory.
	Finding significant improvement we then test the method in more detail
on N--body simulations, considering the problems associated with sampling the
density field  using sparse, magnitude--limited galaxy redshift surveys.
	A method of estimating the magnitude of shot--noise effects on the
deprojected velocity field is derived.
	We find there is a good correlation between the true and  deprojected
density and velocity fields, even in the mildly nonlinear regime, provided that
on large scales galaxies trace the mass. An initial application is made to the
IRAS--QDOT all--sky redshift survey.

\end{abstract}


\section{Introduction}
	In a large number of  cosmological scenarios describing the formation
of structure, gravitational instability is the  principle driving mechanism for
the evolution of stochastic fluctuations in the early universe. Within these
scenarios growth  of the density perturbations on large scales
is initially governed by matter falling into potential wells. One of the
challenges of cosmology has been to relate the observed distribution of
galaxies to these density perturbations.


 Perhaps the most fundamental  limitation is how the galaxy
pattern traces the pattern of density fluctuations. In the standard lore a
linear relationship exists between the moments of the galaxy distribution and
the mass distribution. A new degree of freedom is introduced in the relative
normalization of these fluctuations by a bias parameter, b. This is defined in
the linear  regime by
\be
	\sigma_g = b \sigma_m
			\label{eq: bias}
\ee
where the left hand side is a discrete rms average over  galaxies, and the
right hand side is an rms average over a continuous density  field.
 This definition has found some justification from considering the preferential
 clustering of peaks in random fields, thus enhancing the moments
 of peaks over the field (Kaiser 1984, Politzer and Wise 1984,
 Bardeen \et 1986). In detail this prescription has lead to
 much controversy over how galaxy biasing works and how it may be detected
(Bower \et 1992, Babul \& White 1992).


	A second difficulty arises due the distortion of the density field
when one is observing objects in redshift  projection (Kaiser 1987, McGill
1990, Hamilton 1991). The distortion occurs because of the addition of the
peculiar velocity of a galaxy  to its Hubble recession velocity,  along the
radial line of sight. Instead of having galaxy positions in terms of $(r,
\theta, \phi)$, the position is defined by $(z, \theta, \phi)$, where $z$ is
the measured redshift. This  new set of coordinates is called {\em
redshift--space}. In a high density universe where gravitational  instability
generates peculiar motions, the comoving  displacement of a galaxy due to this
distortion is similar to  that due to intrinsic gravitational clustering.
This further complicates any analysis of the density field
that relies on the 3--dimensional distribution of galaxies.


Yet another difficulty arises due to the nonlinear evolution of the density
field. The scale at which these effects are  important is nominally when
$|\delta({\bf x})|\sim 1$, as  negative fluctuations are bounded by
$\delta({\bf x})>-1$.  As the normalization  of fluctuations in the linear
regime relies on the accurate determination of fluctuations at the present
epoch, one would hope that comparison with the correlation function on large
enough scales would suffice. However it is still not clear exactly how large a
scale is required (Peacock 1991).


	The limited information in measured velocity fields, and  the projected
velocity information in galaxy redshift surveys has  led to new methods of
analysing this data, generically
 called Reconstruction Analysis. This  type of
analysis is interested in recovering information convolved by observational
restrictions, and dynamical evolution. The solutions yield
information
on cosmological parameters and the initial  conditions of clustering.
Precursors of this type of approach can be traced back to van Albada
(1960), Kantowski (1969) and Silk (1974), who used nonlinear gravitational
instability to model the  density and velocity fields of the Virgo
supercluster, assuming spherical symmetry. Peebles (1976) extended this by
solving the  mixed boundary value problem of galaxy motion in Virgo using
linear
gravitation theory, given data on galaxy distances and redshifts.

Bertschinger
\& Dekel (1989) laid the foundations of reconstruction analysis
 by showing that the full
velocity field can be reconstructed from its radial projection by assuming a
potential flow. First the velocity potential is constructed by integrating the
observed radial velocity along the line of sight. Then the
full velocity field is
found by differentiation. From this, the density field can be inferred by the
nonlinear Zeldovich continuity equation (Zeldovich 1970).

Following Peebles (1989, 1990) who solves the
 nonlinear mixed boundary value problem of velocity and position of
galaxies in  the Local
Group subject to the minimisation  of an action principle subject to the
constraint of zero velocities at initial times, Giavalisco \et
(1991) have considered the same approach to the solution to the orbits of
galaxies on large scales, given redshift and distance information. This
solution can also be inverted to yield the velocity field from redshift--space
positions. Yahil (1991)
has recently proposed an alternative to this by formulating a Hamiltonian field
theory for a pressureless self--gravitating fluid.

Reconstruction analysis of the velocity field from redshift--space began with
Davis \& Huchra (1982), who calculated the inferred velocity field about the
Virgo supercluster based on linear theory and the distribution of galaxies in
redshift--space. Following Kaiser (1987), who showed that the galaxy
distribution in redshift--space can be highly distorted, Yahil (1988), Strauss
\& Davis  (1988) and Rowan--Robinson (1991) used the galaxy redshift--space
distribution to infer the  force on each galaxy and iteratively update each
galaxy position until a self--consistent solution of velocities and positions
is found. Kaiser \et (1991) restated this as a linear continuity relation
between density perturbations in real and redshift--space.

Other reconstruction methods
of interest have been introduced by Nusser \& Dekel (1991) use use the
Zeldovich approximation and  the Bernoulli equation to trace potential
fluctuations back to initial conditions.
Weinberg (1992) has also suggested a method for recovering the
primordial density field, based on the monotonic transformation of the
smoothed, present day  density field to Gaussian initial conditions, and Nusser
\et (1991) have suggested transforming from nonlinear density field to linear
density fields to calculate the linear velocity field, based on a
phenomenological relation found in N--body simulations.


	In the present paper, we are concerned with investigating the problem
of  reconstructing the density distribution and velocity field from the galaxy
redshift--space distribution. Our approach  extends
the Kaiser \et (1991) approach of formulating a set of dynamical and continuity
equations to describe the formation and distortion of structure
in redshift--space.
Keeping in mind the three sources of uncertainty --
biasing, redshift--distortion and nonlinearity of the  density field -- we
test our approach on a number of analytic toy models to assess the
basic merits of the method, and then apply it to a set of numerical
simulations. In
addition we shall also be concerned with the effects of sparse sampling and
magnitude limited data on the reconstruction  method. The main, fundamental,
problem with this approach is the  self--consistency of the solution. This can
be achieved by requiring the velocity field obey the dynamical equations of
a self--gravitating field. In order
not to restrict  the analysis to the linear domain, assuming that we have a
well  sampled galaxy distribution, we consider the effects of redshift
projection in the quasi--nonlinear regime. This step is important, as in the
linear regime cosmological parameters enter the solution  in a degenerate
combination, \ie  $g_0=b \Omega_0^{-3/5}$, which is the coupling parameter
between density and
velocity. This can only be independently  determined by going beyond linear
theory.  Hence we wish to solve a set of self--consistent, nonlinear
relations between the velocity and density fields, given a discrete,
redshift--distorted density field as initial boundary condition.
The idealized deprojection procedure developed in this paper is as follows:

	1. The galaxy redshift--space distribution is used to construct  a
smooth redshift--space density field.

	2. The density field is Fourier transformed and the potential
field calculated.

	3. The potential field is used to construct a set of boundary  charges
on the cube to isolated the density distribution from periodic boundary
conditions.

	4. The isolated density field is used to calculate the velocity field
in Fourier space. The velocity field must be a solution to the linear
continuity
equation and dynamical equations of motion.

	5. The velocity field is radially projected and iteratively  used to
solve  the continuity and dynamical equations. Steps 2--5 are repeated until a
stable solution is found.

The rest of this paper is set out as follows. In Section 2 we review the basics
of velocity vector field analysis and redshift--space distortion in linear
theory. In Section 3 we consider our new method of deprojecting  velocity and
density fields from redshift--space, in  quasi-nonlinear theory. We test the
approach on some simple special symmetry toy models. The full analysis is
tested on numerical simulations in Section 4, where we emphasize the problems
associated with discrete, magnitude limited and sparse sampled density fields.
Our conclusions are presented in Section 5.

\section{The Peculiar Velocity Field}

\subsection{Linear Analysis}
\label{subsec-c2.2.3}

	We may specify the positions of particles in an expanding
Friedmann-FLRW cosmology in a reference frame comoving with the  expansion. If
the expansion factor of the universe at a time $t$ is given by $a(t)$, then the
proper position of the particle is
\be
	 \r({\bf x},t) = a(t) \x ,
			\label{eq: pc}
\ee
 where ${\bf x}$ is the particles comoving coordinate.

	Assuming a non--relativistic pressure--free fluid model  for the
material content of the universe, subsequent positions of a particle away from
its initial comoving position  may be described by the Lagrangian mapping,
\be
	 \x({\bf q},t) = \q + \xib(\q,t)
			  \label{eq: gz}
\ee
relating the initial constant time surface, where Eulerian coordinates ${\bf
x}({\bf q},t=0)$ equal the Lagrangian coordinates, ${\bf q}$, to a subsequent
constant $t$ surface by the Lagrangian displacement vector $\xib$. In general
we can decompose $\xib$ into
\be
	\xib= \xib^p + \xib^s,
\ee
where $\xib^p=\nab \Phi$ and $\xib^s=\nab \wedge {\bf A}$ are the  potential
and solenoidal component vectors of the displacement and  $\Phi$ and ${\bf A}$
are potential and potential vector fields respectively.

Differentiating this with respect to time gives us the velocity field in the
comoving frame,
\be
              \dot{\x}({\bf q},t)= \dot{\xib}(\q,t).
							\label{eq: comvel}
\ee
This is related to the peculiar velocity field by $\vb \e \dot{\r}
-(\dot{a}/a) \r = a \dot{\xib}$, which is
the deviation from pure Hubble flow.
Applying the operator $[\nab \wedge]$ to equation
(\ref{eq: comvel}) gives the vorticity vector of the field,
$\omegab \e  a \nab \wedge \dot{\xib} = a \nab \wedge \nab \wedge
\dot{{\bf A}}$. The
vorticity equation in Eulerian coordinates (Peebles 1980) is
\be
	a\dot{\omegab} + (2\dot{a}
	+ \nab. \vb)\omegab -(\omegab. \nab) \vb =0 .
						\label{eq: vort}
\ee
In linear gravitational instability theory there is only an adiabatic decaying
solution to this, $|\omegab| \propto a^{-2}$, in the absence of
non-gravitational
forces. In nonlinear theory a growing mode arises when the vorticity becomes
coupled to the density field, amplifying residual vorticity (Buchert 1992).
 Here we shall
assume that $\omegab(q)$ is zero, and postpone the discussion of vortical
currents around structure for a later paper.

On large scales we expect the mapping, equation (\ref{eq: gz}),
 to coincide with the
results of the linear theory of gravitational instability. The Lagrangian
continuity equation can be linearized
\ba
	\delta 	&=& \frac{1}{det(\delta_{ij} + \de \xi_i / \de q^j)} -1, \nn
		&\ssim& - \nab . \xib, \nn
		&\ssim& -\frac{1}{aH(t)f(t)} \nab . \vb ,
	\label{eq: edum1}
\ea
 where the factor $f(t)$ is the universal growth factor of linear
perturbations,
relative to the Hubble expansion rate, defined as
\be
	f(t) \e  \frac{d \ln \delta(\x,t)}{ d \ln a(t)}.
\ee
Lahav \et (1992) have shown that in general
\be
	f(t_0) = \Omega^{3/5}(t_0)  + \frac{\Omega_{VAC}}{40}
				\left( 1 + \frac{1}{2} \Omega(t_0) \right),
\ee
where $\delta \e \delta \rho/\rho$ is the density contrast and
 where $\Omega^{3/5}$ is the Peebles (1976) approximation
for zero cosmological constant.

Hence the linear velocity field is conveniently found from the Fourier
transform of the continuity equations (\ref{eq: edum1}) giving
\be
	\vb_k = aHf \xib_k = -i a Hf \delta_k \frac{\k}{k^2}.
\ee
This linear relation between the density and velocity fields will be essential
to our analysis.

\subsection{Redshift--space distortions in Linear Theory}

	The projection of phase space onto redshift--space in the comoving
frame is given by \be
	\txb (\xt) = \x +  \idaa (\dot{\xib}(\xt).\hxb) \hxb
	\label{eq: c2.55}
\ee
where we use the tilde notation to represent quantities in redshift--space. The
vector $\hxb$ is the unit radial vector of an observer positioned at the
origin.,
and is invariant under redshift--space transformation.
The amplitude of the distortion effect can be estimated by comparing the ratio
of the Lagrangian displacement vector to  the second term in equation (\ref{eq:
c2.55}), the redshift  displacement vector
\be
	\frac{1}{H} \frac{\left|  \dot{\xib}(\q,t).\hxb \right|}{
		\left| \xib (\q,t) \right|}
		\ssim  |\mu| \Omega^{3/5} \sim 1,
\ee
	where $\mu$ is the cosine of the angle between the local velocity
 vector, $\vb(\q,t)$, and $\hxb$.

	The effect of the distortion on density inhomogeneities can be  easily
illustrated in the linear regime. Consider equation (\ref{eq: c2.55}) as a
mapping between real and redshift--space. This leads to us to the  continuity
expression ({\em cf}. Section \ref{subsec-c2.2.3} for the Lagrangian
continuity equation)
\be
	\trho[\txb(\x,t),t] = \frac{\rho(\xt)}{det(\de \tilde{x}_i/ \de x^j)},
	\label{eq: c2.18}
\ee
	Linearizing we see
\be
	\tdelta (\xt) = \delta - \idaa \nabla . (\dot{\xib}.\hxb) \hxb.
				\label{eq: c2.60}
\ee
	This second term represents the distortion due to displacement
 by the velocity
 field and the effects of changing the coordinate system. The terms
 corresponding to this change in coordinates drops off as $1/x$,
 so the leading term is $\idaa \frac{\de}{\de x} (\dot{\x}. \n)$. Expanding
 this in terms of Fourier components
\be
	\frac{a}{\dot{a}} \frac{\de}{\de x} (\dot{\xib}. \hxb)
		=
		-i f \int d^3k \,
		 \delta_k \mu^2_k e^{-i\k.\x},
\ee
 where $\mu_k= \hat{\k}.\hxb$
	This leads us to the linear relation between the Fourier components
 of the density field in redshift--space and real--space (Kaiser 1987)
 expressed by
\be
	\tdelta_k = \delta_k (1 + f \mu^2_k),
\ee

This indicates that the compressional mode induced by linear infall results in
an amplification of density waves orthogonal to $\hxb$. This enhancement of
features  gives rise to anomalous terms when calculating dipoles and  inferred
velocity fields and so prompted the development of methods to correct for this
distortion (Yahil 1988, Strauss \& Davis 1988, Rowan--Robinson \et 1991, Kaiser
\et 1991, Yahil \et 1991).

Another incentive for tackling the transformation  from redshift to
configuration space is the effect on the  moments of the density field.
Consider first the variance measured over  volume $R_f$.  In redshift--space
this will be amplified by the distortion,
\be
	{\tilde \sigma}^2_0 (R_f) =
		\int dk \, |\delta_k |^2 ( 1+ f \mu_k^2 )^2      =
	A^2  \sigma^2_0 (R_f) ,
 \ee
  where $A$ is usually assumed to be in the range $1 < A < 2$.

	Generally, if we consider the  n--point correlations of the
 field we find
\ba
\lfe{	\langle \tilde{\delta}({\bf x}_1) \tilde{\delta}({\bf x}_2)
  	\cdots \tilde{\delta}({\bf x}_n)	\rangle
		=  }\nn
	&  &
	( 1+ f (\hxb. \nab)^2 (\nabla^2)^{-1})^n
	\langle \delta({\bf x}_1) \delta({\bf x}_2)
  	\cdots \delta({\bf x}_n)	\rangle   ,
\ea
	where we have re--written the Fourier transform of $\mu^2_k$
 as the real--space operator $[(\hxb. \nab)^2 (\nabla^2)^{-1})]$
 (Hamilton 1991). Thus for the case where ${\bf x}_1 = {\bf x}_2 = \cdots
 ={\bf x}_n= {\bf x}$, this reduces to the form:
\be
	\langle \tilde{\delta}^n \rangle = A^n \langle \delta^n \rangle.
\ee
	The constant $A$ can be found by averaging the correlations
 over ${\bf n}$ (Kaiser 1987, McGill 1990, Hamilton 1991).
 This approximately gives $A=(1+\frac{1}{3}f)$.
 Hence, measuring the variance, the redshift projected value could
 be at least a factor of $1.6$ in excess of the underlying field, while the
 skewness may be amplified by a factor of $2$.

	Note that this effect does not jeopardize the
 analysis of the density field when all comparisons
 with theoretical models are made in redshift
 space, eg. Saunders \et  (1991) and Efstathiou \et  (1990),
 but may have the effect of reducing the overall significance of the
 moments of the field.

\section{Quasi-Nonlinear theory of redshift--space distortions}
\label{sec- nonlinear}

\subsection{The Basic Equations}

	The goal of the quasi--nonlinear  theory of distortions is to
 extend the analysis beyond the regime of linear theory and
 into the regime where correlations between galaxies become important,
 but before the point of redshift--space inversion,
 when galaxies occupy multivalued regions in projection.
 This approach may be seen as
 a stepping--stone towards including the constraint of forming structure from
 small perturbations in the early universe, as well as to further understand
 the dynamical process, and its manifestation in the redshift--distortion.
 Finally, via correlation with the true velocity field, measured
 independently, we hope to gain insight into the biasing mechanism
 thought to account for the  relation between galaxy distribution and
 the distribution of mass fluctuations (Kaiser 1984, Politzer \& Wise 1984,
 Bardeen \et 1986).

	In the discussion which follows we shall be concerned
 with the velocity and density fields at the current epoch. Hence,
 for convenience we shall adopt the convention $a_0=\dot{a}_0 =1$.
 The projection equation (\ref{eq: c2.55}) can be written as
\be
	\tx_i(\x,t_0) = x_i + u \hx_i		\label{eq: e 2.67}
\ee
 where
\be
	u(\x,t_0) \equiv  \hx_i v^i (\x,t_0)
\ee
 is the radially projected velocity field.

 The transformation tensor is
\be
	\tJ_{ij} (\x,t_0) = \delta_{ij} + (u\hx_i)_{,j}     \label{eq: trten}
\ee
 where we introduce the comma notation to represent differentiation
 with respect to $\x$. The second term in equation (\ref{eq: trten}) tells
 us how the velocity field distorts the structure in real--space. Expanding
 this term out into parallel and perpendicular contributions
 along the line of sight gives us
\be
	(u\hx_i)_{,j}  = \frac{u}{x} (\delta_{ij} - \hx_i \hx_j)
			+ u' \hx_i \hx_j,
							 \label{eq: ud}
\ee
	where $u'(\x,t_0)=\hx_j \dxj \hx_i v_i(\x,t_0)$ is the gradient of the
 projected velocity field. Equation (\ref{eq: ud}) shows us that the
 relative distortion along the line of sight is solely due to
 the radial gradient of the velocity field. The relative distortion
 also contains a term perpendicular to the radial line due entirely to the
 ratio of the projected velocity to the isotropic Hubble flow.
 This transverse term arises due to fluid elements being projected along a
 conic
 ``funnel'' of constant spherical angle, $d\Omega$, a shift along which
 will produce a squeezing or expansion in the transverse direction.

 Decomposing this further reveals
\be
	(u\hx_i)_{,j}  =  v^m_{\,,j} \hx_m \hx_i +
			\frac{v^m \hx_m}{x} (\delta_{ij} - \hx_i \hx_j) +
			\frac{v^m \hx_j}{x} (\delta_{im} - \hx_i \hx_m),
							\label{eq: decomp}
\ee
	where
\be
	v^j_{\,,i} = \frac{1}{3} \vartheta \delta_{ij} + \sigma_{ij}
			- \omega_{ij},
\ee
	is the usual decomposition of the velocity field  gradients
 into a trace, representing a monopole divergence (volume
 expansion/contraction)
\be
	\frac{1}{3}\vartheta  \equiv v_{i,i}.
\ee
	a trace--free symmetric term (shear)
\be
	\sigma_{ij} \equiv  (v_{(i,j)} -
			   \frac{1}{3} \vartheta \delta_{ij}),
\ee
	and an anti--symmetric part (vorticity)
\be
	\omega_{ij} \equiv v_{[i,j]}.
\ee
	Equation (\ref{eq: decomp}) shows how each velocity gradient
 term is projected radially, while the last two terms correspond to
 the perpendicular distortions in the remaining two directions. This
 also indicates that the distortion tensor is symmetric in its
 indices -- as one expects for a radially projected quantity. This
 symmetry can be fully expressed by writing the transform tensor
 as
\be
	\tJ_{ij}(\x,t_0) = (1+u/x) \delta_{ij}  +
			(u' - u/x) \hx_i \hx_j.
\ee
	From the symmetry properties of this tensor we can easily
 write out the determinant and the inverse tensor as
\ba
	\tJ &\equiv& det \, \tJ_{ij} = (1+u/x)^2(1+u'), \\
	\tJ^{-1}_{ij} &=& \left( \frac{1}{1+u/x} \right) \delta_{ij}
			+ \left( \frac{u'-u/x}{(1+u/x)(1+u')} \right)
			\hx_i \hx_j.
\ea
The continuity equation
\be
	\frac{\Delta \rho}{\rho} = \tJ^{-1} -1
\ee
can similarly be interpreted as the fractional number of galaxies
 ``scattered'' in and out of a volume by their redshift displacement.

\subsection{Field Dynamics}

	So far our set of equations is incomplete until we specify the
dynamics governing the velocity field. The dynamics of the velocity field  are
fully contained in the Euler equations, and self--interaction  takes place
through Poisson's Equation.

	From the discussion of Lagrangian dynamics in Section
\ref{subsec-c2.2.3}, we see that the type of flow is restricted by the
requirement that the mapping we choose coincides with linear theory on
large scales. While we are free to choose any such solution
only a few have been studied in detail; the Zeldovich approximation,
the Frozen Flow analysis of Mataresse \et (1992), and the Lagrangian solutions
of Buchert (1992). As all of these reduce to the linear flow field on large
scales. Indeed the Zeldovich approximation extrapolates the linear velocity
field beyond the linear regime, and may be regarded as the simplest scheme.
We shall adopt this approach and use a linear velocity field
approximation, while concentrating our attention  to solving the
nonlinear continuity equation. Hence the system of equations we shall study
reduces to
\ba
	\delta(\x,t_0) &=& (1+ \tdelta(\tx,t_0))
		(1+u(\x,t_0)/x)^2 (1+u'(\x,t_0)) -1, \nn
 	u(\k,t_0) &\e& \vb(\k,t_0).\hxb = -i f(t_0) \delta(\k,t_0)
				 \frac{\k.\hxb}{k^2}.
\label{eq: e2.81}
\ea
We note that a similar set of equations has been studied by Nusser \et (1991)
for the Zeldovich continuity equation, and a phenomenological universal mapping
from the linear to nonlinear regime found numerically. In the case of the
present set of equations no such solution exists due to the anisotropy of the
redshift distortion.

	In Section \ref{sec-nummeth} we shall discuss our method for solving
these equation on a cubic lattice, and the appropriate boundary conditions.
Before we do so, we first study the distortion effects in more detail, and
analytically test the inversion of equations (\ref{eq: e2.81}) on some
specially symmetric toy models.

\section{Special symmetry toy models}
\label{sec- toys}

\subsection{Plane--wave symmetric solutions}

  Some insight into the nature of the distortion
 can be gained by considering the transformation of a number of toy models
 with special symmetries that allow an analytic treatment. The simplest
 symmetry is one of a plane wave in 1--dimension. Here the potential associated
 with the density field is
\be
	\varphi(x) = - A \cos (kx),
\ee
	where $A$ specifies the wave amplitude and is positive. The peak to
 peak separation is $\lambda = 2 \pi / k $. The real--space density
 field, given by Poisson's equation, reads
\be
	\delta (x) = A k^2 \cos (kx),
\ee
	The observer is situated at the center of one of the ``sheets''. The
 velocity field and its derivative are
\ba
	u(x) &=& - f A k \sin (kx) ,  \label{eq: e47} \\
	u'(x) &=& -f A k^2 \cos (kx),
\ea
 giving us the continuity relation
\be
	\tdelta(\tx)= (A k^2 \cos (kx) +1)(1 - f A k^2 \cos(kx))^{-1} -1.
\ee
	Note that the Jacobian has no transverse term in the 1-d case. In the
 event that $f=-1$, the density field is transformed away. Alternatively,
 a caustic appears in redshift--space if $u'= -f A k^2 \cos(kx) = -1$,
 or at some point when the inequalities $f A k^2 \geq 1$ holds.
 This can be re-expressed as
\be
	\lambda \leq  2 \pi \sqrt{f A},		\label{eq: e41}
\ee
	showing that caustics are more easily generated by short wavelengths
modes. Relation (\ref{eq: e41}) also shows that a necessary condition for
redshift caustic formation in this model is that $f$ must be positive. 	This
simple example illustrates how structure
 can be both created and annihilated by the redshift--space transformation.
 However in this 1--d model, the conditions required for each effect
 are mutually exclusive.

	Using the Zeldovich approximation, which is in fact known to be
 an exact solution to the fully nonlinear equations for the evolution
 of 1--d inhomogeneities (Doroshkevich, 1973), we can extend
 this model and compare the evolved nonlinear density field
 to the nonlinear redshift density pattern. Given that the initial
 potential field is
\be
	\pot (q) = - A \cos (kq),
\ee
 we see that the initial linear velocity field is
\be
	v(q) = \dot{\x}(\zz) = - \dot{D} A k \sin (kq).
\ee
 The function $D$ is the linear growth function for overdensities. Here we
shall
 use it as a time coordinate for convenience.
 The true density field now evolves as
\be
	\delta [x(q,t),D] = (1- D  A k^2 \cos(kq) )^{-1} -1, \label{eq: e44}
\ee
	while in redshift--space the density is seen to be
\be
	\tdelta[\tx(q,t),D] =
		(1- (1+f) D  A k^2 \cos(kq) )^{-1} -1.	\label{eq: e45}
\ee
 The growth of 1--d structures in real--space and redshift--space are plotted
 in Figures \ref{fig: c2.1}a and \ref{fig: c2.1}b, respectively.

	An advantage of the 1-d model is that we can decompose the
density fluctuations into a Fourier series;
\be
	\tdelta(\tx) = -1 + \int d\tx \, \tilde{\rho(k)} \cos (k \tx) ,
\ee
	where the transform of the density field  is
\be
	\tilde{\rho}(k,D) = 2 J_k [k (1+f) D].
\ee

 Figure \ref{fig: c2.2} shows the evolution of the first 100 Fourier modes
up until the time of caustic formation. The caustic forms as a result
of the coordination of all the modes, but note that the first few modes in
fact decrease in amplitude as the caustic forms. The 1--d case of
redshift--distorted fluctuations evolves in a similar fashion to a more
highly evolved density field in the Zeldovich approximation. The
real--space density field looks like the redshift--space density field
after a change of time coordinate, such that
\be
	D(t) \rightarrow (1+f)D(t).
\ee

	Redshift caustics appear when
\be
	 \cos (kq) \geq \frac{1}{A k^2 D (1+f)},
\ee
 at the positions $q = \pm 2 n \pi / k$, and at the time
\be
	D_{\infty} = \frac{1}{ A k^2 (1+f)}
\ee
Substituting $D_{\infty}$ into equation (\ref{eq: e44}) we find that the
density
of the true  field at the time and place of redshift caustic formation is
\be
	\delta_{\infty}  (x,t)  = \frac{1}{f},
\ee
 while the linear density contrast is
\be
	\delta^L_{\infty}  (x,t)  = \frac{1}{(1+f)}.
\ee

	If we alter the model to include the effects of perpendicular
distortions, as in the case of a plane--wave in 3 dimensions, the
redshift--space density field along a line of sight is
\be
	\tdelta[\tx(q,t),D] =
		\frac{(1-D A k^2 j_0(kq))^2}{
		(1- (1+f) D  A k^2 \cos(kq) )} -1.
\ee
The trajectories of the fields in the $\tilde{\delta}(x)$, $\delta(x)$ plane
are plotted in Figure \ref{fig: c2.3a} for the linearized forms  of equations
(\ref{eq: e44}) and  (\ref{eq: e45}), $\delta^L=Dk^2 \pot$, and
$\tdelta^L=(1+f)Dk^2 \pot$, respectively,
 with  coupling $f=0.4$ and wavelength $k^{-1}=100
\Delta x$.  Trajectories travel with increasing $x$ towards the linear
solution. As one  would expect the  dominant feature is the linear relation
$\tdelta^L(x)=\delta^L(x) (1 +f)$. But local density features, near $x=0$
displace the initial  trajectory away from this relation. Thereafter it
steadily converges back to the linear relation.

But if we now plot the true density field against the linear prediction
(Figure \ref{fig: c2.3b}), we see strong deviations, the most pronounced
being in the
negative density regions. The reason for the effects become more clear if
we look at the fully nonlinear case of equation (\ref{eq: e44}) and
 equation (\ref{eq: e45}) shown in Figure \ref{fig: c2.3c}.

The density is now bounded at $\delta=-1$, $\tdelta=-1$, so strong nonlinear
deviations  force the trajectory {\em back } to the true density field. In
the positive quadrant there is no  such constraint and the trajectories diverge
upwards. This reflects the fact that in nonlinear theory, redshift projections
can  strongly perturb the  high density field, but they have a lesser effect on
the low density regions. As with the linear plot, the trajectories near to the
observer are strongly perturbed by the perpendicular distortions, the
strongest effect being near to the origin, where radial lines are close
together. Comparing this nonlinear analysis with the linear  analysis of Figure
\ref{fig: c2.3a} we see that near the origin the two curves obey the same
relation. However as we gradually travel away, the negative side deviates more
from the linear prediction than the positive side, and it continues to do so
until it approaches $\delta=-1$,  $\tdelta=-1$. If the coupling, $f$, is
changed, the  curvature of the  trajectories also changes, becoming more
pronounced for  larger $f$.

	Finally, we present some numerical models of the plane-wave
solution, evolved past the time of redshift--caustic formation.  $10^6$ points
were randomly layed down and each point allocated a velocity  according to its
position, equation (\ref{eq: e47}). The density field was then  evolved
by displacing the points according to the Zeldovich approximation, and the
density calculated. The same prescription was applied to transform each
point onto redshift--space.
	Figure \ref{fig: c2.4} shows one period of
the plane density wave for different
times and coupling. The left hand plots at each era show the redshift
density field (solid line) and the real--space density field (dotted--dashed
line). The right hand plot is in the $\delta$ -- $\tdelta$ plane. At early
time the results of the analytic model are reproduced, but after the
point of formation of a redshift caustic the $\delta$ -- $\tdelta$ plane
splits into two distinct regions, one related to ``voids'', the other to
``clusters''. In the void region, the density in redshift--space is seen
to become nearly constant, with a sharp threshold at the edge of the clusters.
In contrast, the true density smoothly grows from the void centers to the
cluster centers. This effect leads to an asymptotically $\tdelta=$const
line, where the constant is negative. In clusters, the redshift--density
is reduced due to the projection effect smearing out the profile. In
reality we would expect this effect to be smeared out by the effects of
virialization in the cluster centers, and so to be less important. These
strongly nonlinear effects signal the breakdown of the quasi-nonlinear
model. In the real universe, or in simulations, we would expect that
on small enough scales the $\delta$ -- $\tdelta$ plane will become
multivalued.

	Although highly idealized, the plane wave solution  is exact, and a few
quantitative features may be established from it. The distorted field may
deviate substantially from the  linear expectation, especially on small scales.
The negative density regions deviate away from the linear value faster that the
positive density regions, and incline towards the true  density. Also, local
fluctuations can scatter the distorted density field significantly, especially
in high density regions.

\subsection{Spherical top--hat model}

A second, and important, example is the collapse of a spherical  top--hat
perturbation (Peebles 1980), placed at a distance $\x_0$  from the observer.
This model has exact solutions  for both the true flow field  around the
fluctuations, and for the linear velocity field  approximation. This allows us
a first look at how well our approximations can solve the inversion problem.
Its use as a test of models  of clustering dynamics has been discussed by
Nusser \et (1991) and  Materesse \et (1992). We shall present our discussion
along similar lines.

	Consider a spherically--symmetric step--perturbation of radius $R$. The
equation of motion of a shell of proper distance $r$ from  the center is
\be
	\ddot{r} = - \frac{ G M }{r^2},
\ee
	and the first integral gives the energy equation
\be
	\dot{r}^2 = \frac{2 G M }{r} +C_0,
\ee
 where $C_0$ is a constant of integration. These last two equations have
 parametric solutions $r = A (C_k(\theta) -1 ) $,
 $t= Bk(\theta - S_k(\theta))$, where $A^3= G M B^2$, and $\theta$
 is the arc parameter. The parameter $k= - $sgn$(C_0) =\pm1$, depending
 on whether
 the perturbation is overdense or underdense with respect to the
 mean density of the universe. We use the notation $C_k(\theta)$
 for $\cos$ or $\cosh$ when $k=+1$ or $k=-1$, and similarly
 $S_k(\theta)$ for $\sin$ or $\sinh$. We specialize the discussion
 to an $\Omega=1$, $\Lambda=0$, pressureless cosmology for
 simplicity. In this case we find that the evolution of a comoving coordinate
 from its initial Lagrangian position relative to the center of mass
 is
\be
	\x(\q,t) = \frac{1}{2} k \left( \frac{4}{3}\right)^{2/3}
		(1-C_k(\theta))(\theta - S_k(\theta))^{-2/3} \q.
\ee
	The density contrast can now be established from
\be
	\delta(\x,t) = J^{-1}(\x,t) -1,
\ee
 where we impose the condition that $\delta \rightarrow 0$ as
 $t \rightarrow 0$, and find
\be
	\delta(t) = \frac{9}{2} \frac{k (\theta - S_k(\theta))^2}{
			(1-C_k(\theta))^3} -1.
	\label{eq: e67}
\ee
	The velocity field of the top--hat is found by the convective
 derivative of $\x$. This gives us
\be
	\dot{\x}(\x,t) = \left( \frac{3}{2} \frac{k S_k(\theta)
			 (\theta - S_k(\theta))}{(1-C_k(\theta))^2} -1
			\right) \x.
\ee
 If we now place the center of mass of the perturbation at $\x_0$ and
 have the observer at rest with respect to the origin, the projected
 velocity field is
\ba
	u &=&  \left( \frac{3}{2} \frac{k S_k(\theta)
			 (\theta - S_k(\theta))}{(1-C_k(\theta))^2} -1
			\right) (x-\x_0.\x/x) \nn
	  &=&  u_0 (x-\x_0.\x/x).
\ea
	The Jacobian of the redshift--space transformation is
\be
	\tJ(\x,t) =  \left( \frac{3}{2} \frac{k S_k(\theta)
			 (\theta - S_k(\theta))}{(1-C_k(\theta))^2}
			\right) (1+ u_0 (1-\x_0.\x/x^2))^2 ,
\ee
       and the density contrast under this transformation is
\be
	\tdelta_{TH}(\x,t) = \frac{3 (\theta - S_k(\theta))}{
			S_k(\theta) (1-C_k(\theta))}
			(1+ u_0 (1-\x_0.\x/x^2))^{-2}  -1 .
								\label{eq: e63}
\ee
	The surface of vanishing transverse distortion defines the surface of
the redshift caustic. Equation (\ref{eq: e63}) shows that  this is a curved
surface defined by the relation
\be
	(\x -\x_0). \hat{\x}  = 0.
\ee
	As $\x_0$ increases $\x$ and $\x_0$ become nearly parallel and
 the transverse distortion becomes less
 important. For simplicity we shall assume it is negligible and the
 redshift--caustic has zero curvature.

	For overdense regions, $k=1$, we see from equation (\ref{eq: e63}) that
the first redshift--space caustic formation occurs when $S_k(\theta) = 0$,
coinciding with the point of maximum expansion. At this time, the true density
contrast is $\delta_{\infty} =(3 \pi/4)^2-1 = 4.6$, while the linear density
contrast is $\delta^L_{\infty} = 3/5(3 \pi/4)^{2/3} = 1.06$. This compares with
the linear value of $\delta^L=1.69$ for the complete collapse of the object.
This takes place in a characteristic time scale exactly twice that of the time
for the appearance of the first redshift caustic.

	To investigate what effect taking the linear velocity field
 to correct the distortion has on our solution, we want to
 calculate the field under this assumption. In linear theory the
 velocity field and gravitational field are related by
\be
	\vb = \frac{2}{3}  {\bf g}.
\ee
	By transforming from redshift--space to real--space we can calculate
 $\tilde{{\bf g}}$ for a spherically symmetric fluctuation. Focusing attention
on a small patch of space around $\x=\x_0$, only the velocity gradient
terms enter into the distortion. Transforming the integral over density
fluctuations
 introduces the Jacobian, $\tJ$. The inferred gravitational field is found to
be
\be
	\tilde{{\bf g}} = \frac{1}{2} \tJ(\x,t) \tdelta(\x,t) \x,
\ee
	which leads us to
\be
	\tilde{\vb} (\x,t) = -\frac{1}{3} (\delta - u'(\x,t))
				(\x -\x_0)
	\label{eq: e76}
\ee
	for the inferred velocity field around the distorted perturbation.
 This coincides with the result one expects for a linear distortion
 of the density field for a spherically  symmetric perturbation.
 The line of sight  velocity field gradient is
\be
	\tilde{u}' = - \frac{1}{3} ( \delta - u')
		   =  - \frac{1}{3}  \tdelta .
			\label{eq: e69}
\ee
	 Applying this to the linearized redshift continuity equation
$	\tdelta = \delta - u',$
 and  solving iteratively we see that the solution converges to
\be
	\delta^L = \frac{3}{4} \tdelta, \;\;\;\;\;\;\;\;\;\;\;
	 u'     = - \frac{1}{3} \delta^L.
	\label{eq: e122}
\ee
	This is the solution we expect for a linear velocity field  distortion,
where the velocity field is given by equation (\ref{eq: e76}). Using the
nonlinear continuity equation
\be
	\tdelta=(1+\delta)(1+u')^{-1}-1, 	\label{eq: c2.dum1}
\ee
and linear dynamical solution (\ref{eq: e69}), the fields converge towards
\be
	\delta_{QL} = \frac{3\tdelta}{(4 + \tdelta)},  \;\;\;\;\;\;\;\;
	  u'	= - \frac{1}{3} \delta_{QL},
	\label{eq: e82}
\ee
	which is the solution for a nonlinear distortion by a
 linear velocity field.

Figure \ref{fig: c2.5a} compares $\tdelta_{TH}$ (dotted line), given by
equation (\ref{eq: e63}), and the iterated solutions, $\delta_{QL}$ (dot-dashed
line) and $\delta^{L}$ (dashed line) with $\delta_{TH}$ (solid line) for the
spherically--symmetric infall model. As expected in the overdense sector,
 the redshift--distorted overdensity is amplified forming a caustic
 when $\delta_{TH}=4.6$. In the underdense region as for the 1--d
 case, the discrepancy is not so great due to the $\tdelta \geq -1$
 condition. Meanwhile the linear inversion of the distortion is fairly
 poor at recovering the true density, and fails  to be constrained in the
 underdense model. By comparison, the quasi--nonlinear approach
 provides a better fit to the underdense model in real--space and a large
 improvement over the linear for overdensities. In fact the nonlinear
 overcompensate in overdense regions and from the form of equation
 (\ref{eq: e122}) is seen to asymptotically approach $\delta \rightarrow 3$
 for large $\delta_{TH}$. In fact we should not expect the inversion
 method presented here to return any reasonable results beyond
 $\delta_{TH}=4.6$, as once caustic formation takes place in
 redshift--space the transformation equation (\ref{eq: e 2.67}) is no longer
 monotonic and we will no longer be able to algebraically
 invert the continuity equation.

The recovered peculiar velocity field from the linear density deprojection
(equation \ref{eq: e122}) and the nonlinear  deprojected velocity field
(equation \ref{eq: e82}), calculated from equation (\ref{eq: e76}), are  shown
in Figure \ref{fig: c2.5b} (dotted and dashed--dotted lines, respectively)
against the true velocity field around the top--hat model. Again the linear
analysis is only effective over the regime $-0.1 < v < 0.1$, allowing for a $20
\%$ deviation from the  true velocity potential. By comparison, the
quasi-nonlinear model has only a $10 \%$ deviation for $v=0.1$. At $v=\pm 0.3$,
the quasi-nonlinear deviation has increased to $\sim 15 \%$, while the linear
deviation has grown to $33 \%$ $(v=+0.3)$ and $60 \%$ $(v=-0.3)$.

\section{ Numerical Methods}
\label{sec-nummeth}

	Despite the convenience of analytic models, in order to  solve the
continuity and dynamical equations of Section \ref{sec- nonlinear} for a
general density distribution, we must turn to numerical methods. In this
Section we
shall discuss the methods we use to find a stable solution, and test the
method on a number of fully evolved numerical simulations, before going on to
consider the effects of observational restriction on the  method.

	There is no unique way of solving nonlinear differential equations
 and so we are forced to find some numerical approach that will
 satisfactorily fulfill our needs. As we have seen from Section
 \ref{sec- toys},
 the distortion effect is strongly dependent on distance, local
 velocity magnitude, gradient and orientation.
The most hopeful approach for a local transformation to real--space is
to adopt an iterative
 method for determining the roots of nonlinear sets of equations.
 The two most commonly used are the secant and Newton--Raphson
 schemes. The secant scheme has previously been applied by
 Kaiser \et (1991) to the linear inversion problem.
Applying the secant method to the spherical top--hat model gives a quick
convergence to solutions of equations~(\ref{eq: e2.81}).

	Discretization of the set of equations (\ref{eq: e2.81}) gives us the
 form
\ba
	a_{i+1}(\x) &=& A(\x) b_i(\x), \nn
	b_i(\k)     &=& B(\k) a_i(\k),
\ea
	where $i$ is the $i^{th}$ iteration. The density field is sampled by
the grid using the Nearest--Grid--Point (NGP) scheme, and the dynamical
equations solved using the ``poor man's'' potential solver (Hockney \& Eastwood
1988). The linear equations of motion for the fluid are solved using a rapid
elliptic solver  based on the Fast Fourier transform (FFT). While this
algorithm  is not necessarily the most efficient for solving fluid motion
problems, it is useful for systems which also require a substantial  amount of
smoothing  and for isolated source  distributions. We discuss these points in
Section \ref{sec-s6.2}.

	The regular lattice we use here induces and amplifies
 numerical noise due to the resonances of the grid. To dampen
 these unstable oscillations we mixed  the corrected density field
 with a fraction of its previous iteration, in the form
\be
	 \delta_i = \alpha \left\{(1+\tilde{\delta}) \tilde{J}(\delta_{i-1}) -1
			\right\}
	+ \beta \delta_{i-1},
\ee
 where $\alpha + \beta =1$. We find that a ratio $\alpha / \beta =0.25$
 is sufficient to dampen these oscillations.

To improve the probability of convergence to the right root on smaller scales,
we follow Kaiser \et  (1991) and  iterate with a growing coupling constant for
the velocity field, given by
$ 		 f[\Omega(t_0)]b^{-1},$
where $b$ is the linear bias parameter from equation (\ref{eq: bias}), towards
our final value of $b $ and $\Omega_0$. This has the effect of changing the
shape of the function $F(\delta_{i-1})=\delta_i $ -- and its roots -- with each
iteration, the first iteration beginning on a root. The main requirement of
secant is that the function remains approximately linear locally. This can be
maintained if we smooth on large enough scales.

	In Figure \ref{fig: c2.6a} (top plots) we show convergence of the
solution to equations (\ref{eq: e2.81}) (dotted--dashed line) to the simulation
density and velocity potential (solid line), smoothed with a Gaussian window
function of radius $R_f=5 \, h^{-1}$Mpc sampled along an arbitrary skewer,
 for a bias factor and density parameter of unity.
The density field was fully sampled using the complete $64^3$ particles, and
periodic boundary conditions were imposed to coincide with those of the
simulation cube. With Fourier methods, the running time for $\sim 30$
iterations was $\sim 5$ mins. The bottom two plots show the pixel by pixel
comparison of the true density and velocity potentials after convergence and
for the same smoothing length. As indicated by the spherically symmetric
top--hat model, the underdense regions are recovered more accurately.
 The velocity
potential shows slight systematic behaviour at the tails of both of
the  distributions, which we attribute to the tendency of the Zeldovich
approximation to underestimate the correct solution (Nusser \et 1991).

Figure \ref{fig: c2.6b} shows the same configuration, smoothed by a $10 \,
h^{-1}$Mpc Gaussian filter. Again slight systematics can be seen in the
tails.
Finally, we show in Figure \ref{fig: c2.6c} again the same configuration, but
with a smoothing radius of $15 \, h^{-1}$Mpc. The slight bulge in the center of
the comparative distributions (bottom plots) is due to the difficulty in
fitting a zero crossing of the density and velocity potential fields, compared
with the broad peaks of the field.

\section{Realistic surveys}

	We shall now consider the situation in which we do not have full
information about the density field in redshift--space, but only  the density
of {\em galaxies}, which are in some way related to the fluctuations in the
mass--density. We shall assume that galaxies are selected for inclusion
in the survey according to some intrinsic  criteria, based on their apparent
magnitude or size. In addition  to this, we shall also assume  that we do
not have full angular information of the distribution for an observer situated
at the position $\x_0$. The first of these assumptions introduces the  concept
of galaxy biasing, the relationship between density and observed objects. The
second property, that of a selection criteria, results in some  radial
selection function, $\psi(x)$, which, in the case of monotonically decreasing
$\psi(x)$, implies  a cut--off in the depth of the survey. Similarly we can
introduce  an angular selection function, $W_{\Omega}(\theta, \phi)$, for the
probability that a galaxy at that angular position on the sky is included in
the survey. For the   moment we shall assume
that galaxies are a random point process on the  density field, and
mass density, suitably averaged over an ensemble  of realizations.

\subsection{Inclusion of a radial selection function.}

	In the case of a magnitude limited galaxy sample, the observed
 mean local number density drops as a function of distance, as less luminous
 galaxies drop below the flux limit of the survey.
 	The local number density of galaxies observed from $\x_0$ is
\be
	n (\x) 		= \psi (|\x -\x_0|) (1 + \delta(\x)),
\ee
	Inserting this into the derivation of the redshift continuity equation
 where the conserved quantity is now the total number of galaxies,
 we find
\ba
	\Delta (\x) &=& (1 + \tDelta (\x)) (1 + u/x)^2 (1+ u') K(|\x -\x_0|)
			-1, \nn
	K(x) 	    &=&   \frac{\tilde{\psi}(|\x|)}{\psi(|\x|)},
\ea
	where we now define the fluctuation in galaxy number density by
\be
	\Delta (\x) = w(\x) (n (\x) - \lgl n(\x) \rgl),
\ee
 where $n(\x)$ is the discrete summation over galaxies in a finite cell. The
 selection--dependent function, $K(\x)$, corrects for the wrong weighting
 allocated to galaxies as a result of miss-placing the position in
 redshift--space. As both the distortion effect and selection functions
 are functions of radial position, the redshift--distorted quantity,
 $\tilde{\psi}(|\x|)$, is related to its real--space transform by
\be
	\tilde{\psi}(|\x|) = \psi (|\x + u {\bf \hx}|).
\ee

	It is worth commenting on a number of complications regarding the
observer selection function. The first is that in order to calculate the
luminosity function of galaxies, $\Phi(L)$, from which the selection function
is directly calculated, a simple flow model is used to correct the positions
of galaxies from which the luminosity is calculated (Saunders \et  1990). The
major correction is for Virgocentric infall. Applying a number of different
flow models, Saunders \et conclude that the luminosity function is insensitive
to the flow model used, and to the value of $\Omega_0$.

	Secondly, the selection function is calculated from a finite sample of
galaxies, assumed to be a fair sample of the true luminosity distribution of
galaxies. In fact the finite sample will introduce fluctuations about this
distribution. Here we assume that these fluctuations are negilgable.

\subsection{Data sampling and smoothing of the density field }
\label{sec-s6.2}

	In order to implement the scheme described so far, it is necessary
 to assign the density field to a lattice and apply a filter function
 in order to smooth the galaxy distribution. Density assignment uses the
 Nearest--Grid--Point (NGP) scheme (Hockney \& Eastwood 1988), where the
 fluctuation in density
 assigned to a lattice cell is the summation of galaxies in the cell,
 denoted by
\ba
	\Delta (\x)  &=& \int_{\Delta V} d^3x' \,
			\left( \sum_i \delta_D(\x_i-\x') - \psi(|\x-\x_0|)
			\right) w(\x')	 \nn
			& &  + m(\theta_i, \phi_i),
					\label{eq: c2.161}
\ea
	where $w(\x)$ is an arbitrary weighting function,  $m(\theta, \phi)$,
is the mask function (see Section \ref{sec- mask}) and the integral is over the
cell volume. While, in general $w(\x)$ is arbitrary, to let equation (\ref{eq:
c2.161}) coincide with  the density fluctuations we set $w(\x) = 1/\psi(|\x|)$.

In order to reduce the smearing effects due to nonlinear projection, discussed
in Section \ref{sec- toys}, we Gaussian smooth the density field to enforce a
one--to--one mapping between redshift and real--space. 	The size of the
smoothing radius is set by a number of factors. One
factor which we have already considered is that of the nonlinearity of the
density field. From the two simple toy models considered in Section
\ref{sec- toys}, we can take
the overdensities $\delta=4.5$ (top--hat model) or $\delta=1$ (plane--wave) as
indicators of the fully nonlinear regime, when redshift--caustics form.
Taking the linear overdensities at this era, and assuming that the density
field was initially Gaussian distributed, we can compute the fractional volume
of
the density field now in the strongly nonlinear regime. The fractional volume
above the threshold $\nu_{\infty}\equiv \delta_{\infty}/\sigma_0$ is
\be
	V_f (>\nu_{\infty}) = \frac{1}{2} erfc (\nu_{\infty} /\sqrt{2}).
\ee
 If we take the variance on $10 h^{-1}$Mpc to be unity, then about $20 \%$
of the volume has formed a redshift-caustic, assuming $\delta_{\infty} =1$.

	We digress here to note a feature of our  filtering of the density
field in redshift--space. As stated above, there are a number of technical
reasons for wanting to filter the density and   velocity fields. But in
transforming the density field from redshift--space to real--
space, we also
distort the filter function from a sphere to an ellipsoid. While we can hope
that this effect is of negligible importance, it will inevitably place
limitations on the ability of the recovered field to reflect  the true density
field smoothed with spherical filters.

\subsection{Mask functions}
\label{sec- mask}

 Angular selection may be required in a redshift survey for a number of
reasons; the survey may only cover a small fraction of the sky, and so the mask
covers the unobserved regions; at low latitude Galactic emission become
important and a mask may be applied in order to minimise the degree of
contamination to the source distribution; in the case of IRAS galaxies, high
latitude contamination arises due to incompleteness of the IRAS catalog,
and Galactic cirrus confusing the allocation of sources.

	The mask function, $W_{\Omega}(\theta, \phi)$, is an angular selection
function expressing the probability of finding a galaxy at $(\theta,\phi)$.
Translating this onto a lattice gives us the definition  that
$m(\theta, \phi)$ is the fraction of the lattice cell covered by the mask
region. A completely obscured cell has the value  $m(\theta, \phi)=1$, while a
completely unmasked cell has the value $m(\theta, \phi)=0$. This allows us to
extrapolate the local density field into the masked region. In practice this
can lead to anomalously large allocations of density to regions where the
unmasked volume of the cell is small, and the density high. We introduce a
minimum fractional cell volume, below which the density field in the masked
region is set to the mean. A more sophisticated treatment than the one adopted
here would be to normalize the masked region to the surrounding  $n$ cells. An
alternative to using a mask function is to lay down random points in the masked
region (Rowan--Robinson \et 1990, Yahil \et 1991).

\subsection{The Reconstruction method as a boundary value problem}

	As a result of the nonlocal behaviour of equations (\ref{eq: e2.81})
the boundary conditions placed on the lattice take on some importance. In
Section  \ref{sec-nummeth} we were able to use the periodic boundary conditions
of the simulations. For a finite density distribution we have two choices of
boundary condition. In the limit that the density field is a fair sample if the
universe (that is to say its statistical properties are the same as the global
mean values) one can argue that periodic boundary conditions can still be used
and will introduce little distortion to the deprojected velocity field. However
if the sample is not fair the Fourier  modes of the density field, measured
locally,  may not be representative of  their global mean value. Periodic
boundary conditions then extrapolate these modes to all space, resulting in
enhanced velocities and shearing. While the mean velocity can be regarded as a
free parameter, normalised by observed flows, any enhancement of shear can
significantly distort the pattern. Here we prefer to isolate the density field
and only calculate the differential motion of matter arising from local
perturbations. This is equivalent  to a Dirichlet boundary condition for the
velocity potential (or equivalently in linear theory, the density  potential)
where $	\Phi(0) = \Phi(L) =0,$	on the boundary. 	For a lattice
configuration, where the coordinate boundaries coincide with the source
boundaries, the most efficient method  of isolating the source is to  calculate
a  set of screening  ``charges'' on the boundary of the lattice.

\subsection{James' algorithm}

To compensate for the effects of the periodic boundary conditions  imposed by
the Fourier transform of the velocity field, we employ  James' Algorithm (James
1977, Hockney \& Eastwood 1988) to calculate a set of screening ``charges'' on
the boundary of the cube at each iteration. This has  the effect of setting the
potential  outside of the cube exactly to zero. The method is employed by
calculating the potential for a periodic  cube, and then  solving Poisson's
equation on the boundary for zero potential exterior to and on the surface of
the cube. This method can be realised by Fast Fourier transforming the density
field, then calculating the  potential field via the  the relation
\be
	\phi({\bf k},t) = \frac{3}{2}  \Omega(t_0)
			 \frac{\delta({\bf k},t_0)}{ k^2},
\ee
and finally performing the inverse transform only on  the surface of the cube.
This is all the information we need both to solve  the finite difference form
of Poisson's equation, and to determine the  surface charges.  The surface
charges are then Fourier transformed once more on the surface before adding to
the initial  density configuration. This final field is now shielded against
the effects of periodicity by the condition $	\phi({\bf x}) =0 $	for
${\bf x }$ outside the boundary.

\section{Shot--noise velocity statistics}

	In simultaneously solving equations (\ref{eq: e2.81})
 for the peculiar velocity field for
 a finite, noisy, distorted and linear tracer of the density field
 (as we assume galaxies are) we must have some way of determining the
 accuracy of the velocity field at any point. A good approximation
 to the variance of the field at any point is provided by
 calculating the variance of the dipole produced by a spherically
 symmetric, but noisy, galaxy distribution enclosing the
 point. This will remain a good approximation near the center of the
 sample, where the surrounding field is nearly spherical. Closer to
 the edge of the sample this approximation will break down as
 boundary effects begin to dominate.

	In Appendix A we explicitly calculate the expression for
 the variance of a dipole due to Poisson noise near the center of
 a spherically symmetric density field. Here we generalise this result
 to the variance in the velocity field due to shot noise at
 any point ${\bf x}$ from the center of a galaxy survey. We find
 this variance to be
\be
	\sigma^2_{v_{SN}}(x,R) = \kappa \int_V d^3x' \,
		( \langle n({\bf x}') \rangle |{\bf x}' + {\bf x}|^4 )^{-1},
\ee
	where $V$ is the volume of the survey, $\kappa =
 ( f(\Omega_0)b^{-1})^2  / 4 \pi $
 and $\langle n({\bf x}) \rangle = \psi(x)$ is the local
 observable number density of objects for a magnitude
 limited sample. For such a sample
 the selection function $\psi(x)$ is normalized such that
\be
 	\int d^3x \, \psi(x) /M = 1 ,
\ee
where $M$ is the total number of observable galaxies in the survey. A specific
example of the shot noise contribution to the velocity field  can be made from
the selection function determined for IRAS galaxies. Here we pre-empt the
application of the reconstruction method to the QDOT redshift survey (Section
\ref{sec-s10}) and plot the radial selection function in  Figure \ref{fig:
c2.7a} and its inverse, the weighting function allocated to each galaxy as a
function of its position in Figure \ref{fig: c2.7b}. From this selection
function we can calculate the expectation value  of the shot noise component
of a reconstructed density/velocity field from a spherically symmetric density
distribution with a radius $R=150 h^{-1}$Mpc. We plot this function in Figure
\ref{fig: c2.7c}. Given a selection function for the sample, the shot--noise
contribution to the variance of the velocity field can be calculated. This is
plotted in Figure  \ref{fig: c2.7c} as a function of radial position from a
central observer (dotted line) together with the variance of a CDM  average
universe, normalised to $b_8=1.5$ on $8 \, h^{-1}$ Mpc (solid line).  At around
$R=80 \, h^{-1}$Mpc the shot--noise contribution begins to dominate the
statistics.

\section{Numerical results : Testing the Deprojection.}
\label{sec-test}

We now proceed to test the full method on an artificial galaxy catalog
constructed to mimic to IRAS--QDOT redshift survey, Poisson sampled  from the
full CDM simulations. The population is selected to give an effective bias
parameter of $b_I=2.0$  relative to the  underlying density field (see Frenk
\et 1990). The observer is placed at the center of a cube with diameter $L=300
h^{-1}$Mpc, and the distribution flux limited to $0.6$ Jy using the luminosity
function of Saunders \et (1990) for IRAS galaxies.
	The selection function $\psi(x)$ and radial weighting function
$w(x)=\frac{1}{\psi(x)}$ are plotted in Figures \ref{fig: c2.7a}, and
\ref{fig: c2.7b}.
The density field is repeated periodical over this scale to avoid boundary
effects. The observer is again  in the CMBR rest frame. Finally, a mask
function is introduced, which cuts out all  of the galaxies that lie below
$|b|=5^o$, to simulate the effects of Galactic extinction and to test our
method of correcting for this.

We now restate the procedure for the inversion method, described in detail in
earlier Sections, and applied to the series of numerical simulations of the CDM
model:
\begin{enumerate}
	\item As described in Section \ref{sec-s6.2}, to solve the
system of equations \ref{eq: e2.81} self--consistently for the
velocity--density fields,
we put the redshift projected galaxy distribution on a  $32^3$ lattice. The
galaxies are restricted to lie within a spherical volume of radius $R=150
h^{-1}$Mpc.
	\item Each grid point is allocated a density, defined by the
surrounding galaxies using the NGP scheme described by equation
 (\ref{eq: c2.161}). The expectation value of the density is set
by the observer
selection function, $\langle n_i \rangle = \psi(x_i)$, where $x_i$ is the
radial distance from the observer to the  grid point. Averaging over each grid
cell after the allocation, and correcting for the masked regions, we find
typically that $98 \%$ of the total mass of the survey has been recovered. The
lost $2 \%$ due to approximations in calculating the masked volume, and
fluctuations in the selection function of the simulation.
	\item  The density field  is then smoothed with a Gaussian filter to
eliminate cusps and any folding of the density field, as well as to alleviate
discreteness effects.
	\item A set of boundary  charges is calculated using the James
Algorithm.
	\item The velocity and density  fields are then iterated towards the
roots of equation (\ref{eq: e2.81}), via a secant method.
\end{enumerate}

	Figures \ref{fig: c2.8} shows an  arbitrary slice
through two deprojected realizations of the  CDM model catalog with a smoothing
length of $5 \, h^{-1}$Mpc (top panel) and
$10\, h^{-1}$Mpc (bottom panel). The density
field is represented by a continuous greyscale with the corresponding true
velocity field superimposed as a vector field.  The initial galaxy distribution
was projected onto redshift--space  and subsequently smoothed and the true
density and the velocity fields deconvolved.
Figure \ref{fig: c2.8b} shows two more slices, this time with smoothing lengths
of $10 \, h^{-1}$Mpc (top panel) and
$20\, h^{-1}$Mpc (bottom panel).

Below a filter radius of $5 h^{-1}$Mpc the linearity of the velocity
field is not strictly applicable in the simulations, and the velocity and
acceleration vectors are no longer aligned. Davis \et (1991) have shown this
effect on scales of about $5 h^{-1}$Mpc. On scales below $5 h^{-1}$Mpc   the
effects of multistreaming after orbit crossing also become  important, and such
effects cannot be followed by the Zeldovich  approximation (eg. Nusser \et
1991). The choice of our filter radius is also dictated by the sparseness of
the galaxy distribution on large scales, which is dominated by shot noise
effects.
We note that the general features of the flow are in good agreement  with the
density field away from the boundary (as one would hope, overdensities and
underdensities occur in roughly the same region, with  mass flowing into  large
overdensities and out of the underdense regions).

	Figure \ref{fig: c2.9} plots the deprojected density field against  the
true density field for smoothing lengths of $10 \, h^{-1}$Mpc (top panel) and
$20\, h^{-1}$Mpc (bottom panel), showing that there is  strong agreement
between the two. Davis \et (1991) have used numerical simulations  to assess
the effects of a discrete, magnitude limited sampling of the  density field.
For a filter length of $1000$ kms$^{-1}$, they find a scatter about the mean
$\delta({\bf x})$ of $\sim  0.05 $. After deprojection we find  that this
scatter has increased by  about a factor of  2. In the case of filtering on a
scale of $2000$ kms$^{-1}$ we find an overall scatter of  0.05. Davis \et  also
find a degree of curvature in the density relations due to random
offsets in the
observer selection function. In our plot there is no scatter about the
normalization, since both the original simulation and our deprojection use
identical  selection functions. Any curvature present is thus entirely  due to
redshift distortion.

	To assess the scatter produced in the velocity field
due to the sum total of selection effects, Poisson noise, and boundary
conditions, we plot in Figures \ref{fig: c2.11a}  the
simulation 1-d  velocity field in the $x$--direction against that
of the reconstructed
velocity field, smoothed on a scale of $5 \, h^{-1}$Mpc. The plot shows
a random subsample of all the grid pixels. Although noisy, there is clearly
a correlation between the fields.

Figure \ref{fig: c2.11b} shows the same field for a smoothing length of
$10 \, h^{-1}$Mpc. Here we see a better correlation between fields, but with
slight systematic effects appearing in the tails of the plot.  This is most
probably due to
limitations of the Zeldovich approximation, which tends to overestimate the
correction required.

\section{Reconstructing the Density Maps of QDOT}
\label{sec-s10}

	Having tested our reconstruction method, we apply it to the IRAS--QDOT
all-sky redshift survey (Rowan--Robinson \et 1990, Lawrence \et 1992).
 Figures \ref{fig: c2.12},  \ref{fig: c2.13} and  \ref{fig: c2.14}
 show  slices through the z--axis at $z=130 \, h^{-1}$Mpc,
$z=110 \, h^{-1}$Mpc,
and $z=90 \, h^{-1}$Mpc, respectively, where the
coordinate system is define so the Local group lies at the origin, and
the x,y,z axes lie along the directions $(l,b)= (0,0)$, $(l,b)= (90,0)$,
$(l,b)= (0,90)$ in galactic coordinates. Clearly shown are the Hercules
supercluster ($x=70\, h^{-1}$Mpc, $y=5 \, h^{-1}$Mpc), and Bootes void
($x=5\, h^{-1}$Mpc, $y=80 \, h^{-1}$Mpc) and the respective flow fields.

\section{Summary and Conclusions}

The transformation from the density field to the velocity field is not
straightforward, as the initial density field has undergone  substantial
evolution and is now nonlinear on moderate scales. Also the density field is
viewed in projection in redshift--space. This has the effect of amplifying
nonlinearities in an anisotropic manner.

	In this paper we have constructed a general method, based on the
gravitational instability scenario for extracting information from raw
redshift--space distance measurements. This allows the  deprojection of the
true density field, and the true  velocity field from redshift--space density
fields. This is  achieved in a physically plausible and self--consistent manner
based on an equation of continuity for  the  local transformation of fluid
elements in redshift--space.  The basic approach was tested on two extreme
analytic models of structure formation, namely the pancake model and the
spherical top--hat model. This allows a clear comparison between exact
solutions and the reconstruction method used. We find that retaining a linear
velocity field, but using the exact redshift--space continuity equation leads
to substantial improvement over linear theory, provided that the mapping
between redshift--space and real--space is one--to--one. In high density peaks,
the local velocity field smears out the redshift--space density, even without
the strongly nonlinear effects of virialisation.

	The full numerical method for solving the distortion equations was
tested on numerical simulations of a standard CDM universe to test the
reconstruction on a general density field. Using the secant iterative scheme
and the FFT to smooth and calculate the potential we find that the density and
velocity fields can be accurately recovered. Slight systematic disagreement
appears at the extreme range of the fields which we attribute to limitations of
requiring a linear velocity field.

Observationally the density field must be derived from the galaxy density
distribution, introducing sampling and discreteness effects. Nonuniform radial
sampling of the density field is compensated for by the use of a mask function,
where the density of the masked volume is set equal to the locally cell
density. The appropriate boundary value to use for a finite density field is
discussed, and the choice of zero potential on the boundary is made in
preference to periodic boundary conditions. This choice is made in order to
reduce the distortion effects in the velocity field from periodicity. James'
(1977) Algorithm is used to screen the lattice.  An approximation to the local
shot noise contribution to the velocity field is derived, assuming Poisson
sampling of the density field and in the absence of smoothing. We are currently
working on a method of dealing with the statistical  noise introduced by galaxy
discreteness. Here we smoothed density field to eliminate discreteness, smooth
over masked regions where the extrapolation method breaks down, and to recover
a linear velocity field. Smoothing is also required to allow the use of a
transformation equation, as the inverse relation must exist. This breaks down
at the time of caustic formation. Based on analytic models this occurs when the
linear overdensity is in the regime $0.5 \ls \delta^L_{\infty} \ls 1.06$. We
deduce that $R_f \gs 5 \, h^{-1}$Mpc is required to meet these criteria.

	The method so far described has been tested on a set of simulated
galaxy catalogs to test the effects of sparse sampling, masking and finite
density field. Although  the noise level is increased substantially, we still
find a strong correlation with the true fields.	Systematic sources of error,
which are not modeled here, may also affect our results. Errors in the
selection function  of the observer may lead to an incorrect estimate of
densities with increasing depth. Such a  problem could arise from errors in the
galaxy  flux assignments in the calculation of the luminosity function. The
redshifts are also subject to errors, but these are mainly much smaller than
other quantifiable errors. Another important issue which  we have not touched
upon is the appropriate reference frame from which to minimize  relative errors
in the density and velocity  fields. It has been argued elsewhere (Kaiser \et
1991, Davis \et  1991) that the deprojection in the observational reference
frame will inhibit  the amplification of these relative errors.  In the case of
real  redshift data, this will be the Local Group rest frame. We shall discuss
this and other questions relating to the statistical information stored in the
reconstruction method, and how to eliminate rest frame dependancey  in a later
paper (Taylor \& Rowan-Robinson 1992).

	At a more fundamental level there are questions relating to  biasing
and the density parameter of the universe. The bias parameter is closely
related to the poorly constrained  theories of galaxy formation. The
methods employed here do not break the linear $f(\Omega_0)/b$ degeneracy, and
so cannot distinguish between a low density universe  with weak biasing, and a
high density universe with strong biasing (so long as the ratio is the same).
This  degeneracy can be broken by using the Lagrangian method to transform
density maps back to initial conditions, as well as correcting the
redshift--space distortion. Using the Zeldovich approximation, the degeneracy
can be broken by going to quasi-nonlinear theory in the real and
redshift--space density fields, while keeping the velocity field linear.
Work on this method is currently in progress.
Such a break would allow one to determine both the  universal growth
function, and spatial variations in the biasing parameter.

 	The application of these methods to quantifying the flow
 field will be discussed in a second paper, and applied to the IRAS--QDOT
 redshift survey (Taylor and Rowan--Robinson 1992).

\section*{Acknowledgements}

	ANT acknowledges the receipt of a SERC Postgraduate
 research grant. We thank Carlos Frenk for useful discussions and for the
 generous provision of his N--body simulations of the IRAS catalog.
 We also thank  Will Saunders, George Efstathiou,
 and Marc Davis for useful discussions, Blane Little, Peter Coles
 and an anonymous referee
 for helpful suggestions and comments on an earlier draft of this paper.

\section*{References}

\nd  Bardeen, J. M., Bond, J. R., Kaiser, N. \& Szalay, A. S., 1986, {\apj},
 {\bf 304}, 15.

\nd Bertschinger, E., and Dekel, A., 1989, {\apj}, {\bf 336}, L5.

\nd Babul, A., \& White, S., D., M., 1991, \mn, {\bf 253}, 437.

\nd Bower, R., G., \et 1992, \mn, in press.

\nd Buchert, T., 1992, \mn, {\bf 254}, 729.

\nd Davis, M., \& Huchra, J., 1982, {\apj}, {\bf 254}, 437.

\nd Davis, M., Strauss, M., A. \& Yahil, A., 1991, {\apj}, {\bf 372}, 394.

\nd Doroshkevich, A., G., 1973, {\sl Astrophysical Lett.}, {\bf 14}, 11.

\nd Efstathiou, G., Kaiser, N., Saunders, W., Lawrence, A., Rowan--Robinson,
 M.,  \\
Ellis, R., S. and Frenk, C., S., 1990, {\mn}, {\bf 247}, 10p.

\nd Frenk, C., S., White, S., D., M., Efstathiou, G. and Davis, M.,
 1990, {\apj}, {\bf 351}, 10.

\nd Giavalisco, M., Mancinelli, B. Mancinelli, P., and Yahil, A., 1991,
 preprint.

\nd Hockney, R., W. \& Eastwood, J., W., 1988, {\sl `Computer Simulations using
Particles'}, Adam Hilger, Bristol.

\nd Hamilton, A., J., S., 1991, preprint.

\nd James, R., A., 1977, {\sl J. Comp. Phys.}, {\bf 25}, 71.

\nd Kaiser, N., 1984, {\apj}, {\bf 284}, L9.

\nd Kaiser, N., 1987, {\mn}, {\bf 227}, 1.

\nd Kaiser, N., Efstathiou, G.,  Ellis, R., S., Frenk, C., S., Lawrence, A.,\\
 Rowan--Robinson, M. and Saunders, W., 1991, {\mn}, {\bf 252}, 1.

\nd Kantowski, R., 1969, \apj, {\bf 155}, 1023.

\nd Lahav, O., Lilje, P., B., Primack, J., R. and Rees, M., J.,
    1991, {\mn}, {\bf 251}, 128.

\nd Lawrence, A., \et, 1992, in preperation.

\nd McGill, C., 1990, {\mn}, {\bf 242}, 428.

\nd Nusser, A., Dekel, A., Bertschinger, E. and Blumenthal, G., R., 1991,
 {\apj}, {\bf 379}, 6.

\nd Nusser, A. \& Dekel, A. 1992, \apj, in press.

\nd Peacock, J., A., 1991, {\mn}, {\bf 253}, 1p.

\nd Peebles, J., A., 1976, \apj, {\bf 147}, 859.

\nd Peebles, J., P., E., 1980, {\sl The Large Scale Structure of the
 Universe}, (Princeton: Princeton\\
\spss    University Press).

\nd Peebles, J., P., E., 1989, \apjl, {\bf 344}, L53.

\nd Peebles, J., P., E., 1990, \apj, {\bf 362}, 1.

\nd Politzer, H., D., and Wise, M., B., 1984, {\apj}, {\bf 285}, L1

\nd Rowan--Robinson, M., Lawrence, A., Saunders, W., Crawford, J.,
       	Ellis, R., S., Frenk, C., S., Parry, I., Xia, X.--Y.,
   	Allington--Smith, J.,  Efstathiou, G.
       	and Kaiser, N., 1990, {\mn}, {\bf 247}, 1.

\nd Saunders, W., Rowan--Robinson, M., Lawrence, A., Efstathiou, G.,
       Kaiser, N., Ellis, R.and  Frenk, C., S., 1990, {\mn},
       {\bf 242}, 318.

\nd Saunders, W., Frenk, C., S., Rowan--Robinson, M.,  Efstathiou, G.,
   Lawrence, A., Kaiser, N.,
        Ellis, R., S, Crawford, J.,
          Xia, X.--Y. and Parry, I.,
         1991, {\na}, {\bf 349}, 32.

\nd Silk, J. 1974, \apj, {\bf 194}, 215.

\nd Strauss, M. and Davis, M., 1988, {\sl Large--Scale Motions
    	in the Universe : A Vatican \\
\spss	Study Week.}, (Princeton Univerity Press : Princeton).

\nd Taylor, A., N. and Rowan-Robinson, M., 1992, in preparation.

\nd van Albada, G. B., 1960, {\sl Bull. Ast. Inst. Neth.}, {\bf 15}, 165.

\nd Weinberg, D., 1992, in preperation.

\nd Yahil, A., 1988, {\sl Large--Scale Motions in the Universe :
        A Vatican Study \\
\spss	Week.}, (Princeton University Press : Princeton).

\nd Yahil, A., 1991, preprint.

\nd Yahil, A., Strauss, M., A., Davis, M. and Huchra, J., P., 1991,
  {\apj}, {\bf 372}, 380.

\nd Zeldovich, Y., B., 1970, {\asap}, {\bf 5}, 84.

\newpage

\section*{ Appendix A}

\subsection*{Variance Of The Velocity Field due to Discreteness Noise.}
	When integrating over the density field to calculate the
 velocity at a point, noise is introduced into the calculation
 by the discreteness of the observed galaxy distribution. For a
 magnitude limited survey, this noise is an increasing  function
 of distance from the observer. The variance of the velocity field
 near to the center of the survey can be approximated by the
 variance in the dipole due to noise in the survey.

	We proceed by calculating the variance of the dipole
 for an observer positioned at the center of a finite but spherically
 symmetric galaxy distribution. For this, the simplest approach is
 to find the characteristic
 functional for the velocity field. The induced velocity produced by
 a finite cell of volume $\Delta V$  at  a distance ${\bf r}$ from the
 origin is
\be
	{\bf v}({\bf x})= \Delta V
       \left( \frac{N({\bf x})-\langle N({\bf x}) \rangle}
	{\langle N({\bf x}) \rangle} \right)
       	\frac{{\bf x}}{x^3},
\ee
 where, for an observer limited by a flux threshold, $\langle N({\bf x})
 \rangle =\psi(x)\Delta V$ is the total number of galaxies expected
 in a cell, and
  $N({\bf x}) = \int d^3x \sum_i \delta_D({\bf x} -{\bf x}_i)$ is the
 number found in the cell. We have ignored numerical factors.
 The observer selection function $\psi(x)$ is normalised so that
\be
 	\int_V d^3x \, \psi(x)/M = 1.
\ee
 where $V$ is the survey volume, and $M$ is the number of observed galaxies.

 	The contribution to the characteristic functional from this cell
  is then
\ba
	G_1[\mbox{\boldmath $\mu$}({\bf x})]& =&
		\langle \exp (i \mbox{\boldmath $\mu$}({\bf x})
			{\bf . v}({\bf x})) \rangle  \\
		& =&  \int d^3x \,  d\delta \, \sum_n  \,
		  p_1(n({\bf x}),\delta({\bf x}),{\bf x})
		 e^{i\mbox{\boldmath $\mu$}({\bf x}){\bf . v}({\bf x})},
\ea
 where angled brackets indicate averaging over all
 random variables in the expression, and
 $j_n(x)$ is a spherical Bessel function of order $n$. The
 probability term can be decomposed into
 $p(n({\bf x})|\delta({\bf x}),{\bf x}) \,
  p(\delta({\bf x})|{\bf x}) \, p({\bf x})$.
 For the
 present we will determine the purely shot noise component, for
 $\delta=0$. The distribution function for a randomly placed cell
 in a spherical volume is given by $p(x)=1/V$.

  If we keep the cells sizes $\sim x_0^3$, where $x_0$ is the
 correlation length for galaxy clustering ($x_0 \simeq 5 h^{-1}$Mpc)
 we can assume each cell is statistically independent. The desired
 functional  then simplifies to
\ba
	G[\mu]    & = &  \prod_{\forall cells} G_1[\mu]\, \, =  \, \,
			 G_1[\mu]^{(V / \Delta V)},  \\
	          & \simeq &  \langle e^{(\alpha(\mu))} \rangle,
\ea
	where
 \[	\alpha(\mu) = 4 \pi  \int_V dx \,    x^2
		(j_0(\mu({\bf x}) v({\bf x}))  - 1)/V.
 \]

	The shot noise variance found at the center of a spherically
 symmetric point distribution is then
\ba
 \sigma^2_{v_{sn}} & = & \left[ \frac{\delta^2 }{\delta \mu^2} G[\mu]
						 \right]_{\mu=0}\\
  	    & = & \int_V dx \, ( x^2 \langle n({\bf x})\rangle  )^{-1}.
\ea

\newpage

\section*{Figure Captions}
\bfg
\label{fig: c2.1}
The evolution of a half--period density wave in (a) real--space and (b) in
redshift--projected space. The lower bound of $\tdelta \geq -1$ for
fluctuations restricts the growth and projection of negative fluctuations.
Positive fluctuations are not restricted, and the redshift--space projection
evolves like a more highly evolved density field.
\efg

\bfg
\label{fig: c2.2}
Spectral analysis of the redshift--space 1-d density field, showing the
evolution of the first 100 Fourier modes. The caustic is generated as a result
of the coordination  of modes. Note that the first few modes decrease from the
linear expectation as the $\tdelta > -1$ is reached.
\efg

\bfg
\label{fig: c2.3a}
One dimensional trajectory for increasing $x$ in the true density field,
$\delta(x)$,  and redshift projected density field $\tilde{\delta}(x)$ plane,
for the linearized  plane--wave solution.
\efg

\bfg
\label{fig: c2.3b}
Linear theory redshift density contrast, $\tdelta^L$ plotted against the
true density contrast $\delta$. The true density field is constrained by
$\delta >-1$, distorting the trajectory away from its linear value.
\efg

\bfg
\label{fig: c2.3c}
		Fully nonlinear redshift--space and real--space density fields.
\efg

\bfg
\label{fig: c2.4}
Numerical model of the highly nonlinear evolution of the redshift density field
past redshift--caustic formation. Left hand panel shows the true and
redshift--distorted density field profiles. The right hand panel shows the
evolution in the $\delta$ -- $\tdelta$ plane. In voids $\tdelta$ tends towards
a constant negative value, while in clusters the density profile is smeared as
a result of redshift--inversion. Here galaxy positions are superimposed.
\efg

\bfg
\label{fig: c2.5a}
Analytic top--hat model of an evolving cluster/void. The deprojection methods
can be modeled analytically for a comparative study of the approximations. Here
we show the deviation of the linear (dotted line) and quasi-nonlinear
(dotted-dashed line) deprojection from the true density field (solid line). The
dashed line is the redshift--space density contrast.
\efg

\bfg
\label{fig: c2.5b}
Analytic top--hat model of an evolving cluster/void. The deprojection methods
can be modeled analytically for a comparative study of the approximations. Here
we shows  the velocity profile from the linear (dotted line) and
quasi-nonlinear
(dotted-dashed line) deprojections, compared with the true value (solid line).
\efg

\bfg
\label{fig: c2.6a}
Numerical reconstruction of simulations of a CDM cosmology: smoothing length of
$5\, h^{-1}$Mpc.
(a) Top plot: Starting with the redshift density field and a high
 bias, the redshift continuity equation is solved iteratively towards the
 true density (left plot) and velocity
(right plot) fields. (b) bottom plot: pixel by pixel comparison of the density
and velocity potential fields.
\efg

\bfg
\label{fig: c2.6b}
Numerical reconstruction of simulations of a CDM cosmology: Same as Figure
\ref{fig: c2.6a} with smoothing length of $10\, h^{-1}$Mpc.
\efg

\bfg
\label{fig: c2.6c}
Numerical reconstruction of simulations of a CDM cosmology: Same as Figure
\ref{fig: c2.6a} with smoothing length of $15\, h^{-1}$Mpc.
\efg

\bfg
\label{fig: c2.7a}
Radial selection function, $\psi(x)$ for IRAS galaxies, normalised to the
IRAS--QDOT redshift survey.
\efg

\bfg
\label{fig: c2.7b}
Radial weighting function, $w(x)=1/\psi(x)$ for IRAS--QDOT galaxies. The radial
weighting function is equal to the cube of the average observed galaxy
separation.
\efg

\bfg
\label{fig: c2.7c}
Shot--noise variance in velocity associated with each grid point as a function
of radial distance from the observer.
\efg

\bfg
\label{fig: c2.8}
	Greyscale maps of the density field with  the velocity vector
field overlaid, and smoothed on scales of $5 \, h^{-1}$Mpc (top panel) and
$10\, h^{-1}$Mpc (bottom panel). Both maps are a random slice through different
realisation of a  CDM simulation flux limited galaxy distribution. The galaxy
distribution has been distorted by redshift--projection, and subsequently
deprojected.  The  galaxies randomly sample the density field and are flux
limited using the selection function  given by Saunders \et (1990).
\efg

\bfg
\label{fig: c2.8b}
	Greyscale maps of the density field with  the velocity vector field
overlaid, and smoothed on scales of $10 \, h^{-1}$Mpc (top panel) and
$20\, h^{-1}$Mpc (bottom panel).
  Both maps are a random slice through
different realisation of a  CDM simulation flux limited galaxy distribution.
The galaxy distribution has been distorted by redshift--projection, and
subsequently deprojected.  The  galaxies randomly sample the density field and
are flux limited using the selection function  given by Saunders \et (1990).
\efg

\bfg
\label{fig: c2.9}
	Pixel by pixel comparison of the true density field and the
reconstructed density field for smoothing lengths of $10 \, h^{-1}$Mpc (top
panel) and  $20\, h^{-1}$Mpc (bottom panel).
\efg

\bfg
\label{fig: c2.11a}
True velocity at each grid point plotted against  the recovered velocity field.
Only the x--direction is  shown because of the isotropy of the flow.    The
filtering is on scales of $R_f=5 h^{-1}$Mpc.  This plot shows a rough
correlation between the  predicted velocity flow and the true velocity flow.
The
scatter  is mainly due to the inaccurate calculation of the flow field near
the edges of the  galaxy distribution and near those regions which are masked,
and the considerable shot noise contribution from large radii.
\efg

\bfg
\label{fig: c2.11b}
True velocity at each grid point plotted against  the recovered velocity field.
Only the x--direction is  shown because of the isotropy of the flow.    The
filtering is on scales of $R_f=10 h^{-1}$Mpc.  This plot shows an improved
correlation between the  predicted velocity flow and the true velocity flow.
\efg

\bfg
\label{fig: c2.12}
	Slice through the $z=130 \, h^{-1}$Mpc plane of the QDOT survey
for a smoothing length of $20\, h^{-1}$Mpc (see text for definition of
coordinate system). Clearly shown are the Hercules
supercluster ($x=70\, h^{-1}$Mpc, $y=5 \, h^{-1}$Mpc), and Bootes void
($x=5\, h^{-1}$Mpc, $y=80 \, h^{-1}$Mpc) and the respective flow fields.
\efg

\bfg
\label{fig: c2.13}
	Slice through the $z=112 \, h^{-1}$Mpc plane of the QDOT survey
for a smoothing length of $20\, h^{-1}$Mpc (see text for definition of
coordinate system).
\efg

\bfg
\label{fig: c2.14}
	Slice through the $z=90 \, h^{-1}$Mpc plane of the QDOT survey
for a smoothing length of $20\, h^{-1}$Mpc (see text for definition of
coordinate system).
\efg

\end{document}